\documentclass[ALICE,manyauthors]{cernphprep}

\usepackage[comma,square,numbers,sort&compress]{natbib}
\usepackage{hyperref}
\usepackage{lineno}
\usepackage{bm}

\begin{document}%

\begin{titlepage}
\PHyear{2018}
\PHnumber{101}      
\PHdate{10 May}  
%

\title{Inclusive J/$\psi$ production \\ at forward and backward rapidity in p--Pb collisions at \boldmath${\sqrt{{\bf{s}}_{\rm \bf{NN}}}}$ = 8.16~TeV}
\ShortTitle{Inclusive J/$\psi$ production in p--Pb collisions at ${\sqrt{s_{{\rm NN}}} = 8.16}$~TeV}   

\Collaboration{ALICE Collaboration\thanks{See Appendix~\ref{app:collab} for the list of collaboration members}}
\ShortAuthor{ALICE Collaboration} 

\begin{abstract}
Inclusive J/$\psi$ production is studied in \mbox{p--Pb} interactions at a centre-of-mass energy per nucleon--nucleon collision $\sqrt{s_{\rm NN}}=8.16$ TeV, using the ALICE detector at the CERN LHC. 
The J/$\psi$ meson is reconstructed, via its decay to a muon pair, in the centre-of-mass rapidity intervals $2.03<y_{\rm {cms}}<3.53$ and $-4.46<y_{\rm {cms}}<-2.96$, where positive and negative $y_{\rm {cms}}$ refer to the p-going and Pb-going direction, respectively. The transverse momentum coverage is $p_{\rm T}<20$ GeV/$c$. In this paper, $y_{\rm cms}$- and $p_{\rm T}$-differential cross sections  for inclusive J/$\psi$ production are presented, and the corresponding nuclear modification factors $R_{\rm pPb}$ are shown. Forward results show a suppression of the J/$\psi$ yield with respect to pp collisions, concentrated in the region $p_{\rm T}\lesssim 5$ GeV/$c$. At  backward rapidity no significant suppression is observed.
The results are compared to previous measurements by ALICE in \mbox{p--Pb} collisions at $\sqrt{s_{\rm NN}}=5.02$~TeV and to theoretical calculations. Finally, the ratios $R_{\rm FB}$ between forward- and backward-$y_{\rm {cms}}$ $R_{\rm pPb}$ values are shown and discussed. 

\end{abstract}
\end{titlepage}
\setcounter{page}{2}

%
%
\section{Introduction}

Quarkonium production in nuclear collisions is sensitive to the temperature of the produced medium. In particular, the various quarkonium states are expected to melt in a Quark-Gluon Plasma (QGP), due to screening of the colour interaction in a deconfined state~\cite{Matsui:1986dk}. In addition, the abundant charm-quark production in the multi-TeV collision-energy range can also lead to a (re)generation of charmonia during the QGP evolution and at the phase boundary~\cite{BraunMunzinger:2000px,Thews:2000rj}. A detailed investigation of these processes was carried out by ALICE, which has measured inclusive J/$\psi$ production in \mbox{Pb--Pb} collisions down to zero transverse momentum ($p_{\rm T}$). These results were reported at centre-of-mass energies per nucleon pair $\sqrt{s_{\rm NN}}=2.76$ and 5.02 TeV, at forward centre-of-mass rapidity $y_{\rm {cms}}$~\cite{Abelev:2012rv,Abelev:2013ila,Adam:2015isa,Adam:2016rdg} for both energies, and at central $y_{\rm {cms}}$ for $\sqrt{s_{\rm NN}}=2.76$ TeV~\cite{Adam:2015rba}.
The nuclear modification factor $R_{\rm AA}$ was evaluated, corresponding to the ratio between the \mbox{Pb--Pb} and the pp production cross sections, normalised to the number of nucleon-nucleon collisions. A suppression of the J/$\psi$ was observed, as indicated by values of $R_{\rm AA}$ smaller than unity. However, the suppression was found to be systematically smaller with respect to results obtained at RHIC energies~\cite{Adare:2011yf,Adamczyk:2016srz}. In addition, the suppression effects were less strong at low $p_{\rm T}$. These observations, together with the comparison to theoretical model calculations~\cite{Zhao:2011cv,Zhou:2014kka,Andronic:2012dm,Ferreiro:2012rq} and the measurement of a non-zero elliptic flow for the J/$\psi$~\cite{Acharya:2017tgv}, imply that a fraction of the J/$\psi$ yield is produced via recombination of charm quarks, and that recombination is more prevalent at low $p_{\rm T}$, where the bulk of charm-quark production occurs.

In addition to effects connected with the hot medium, cold nuclear matter (CNM)  effects are expected to influence the charmonium yield in nuclear collisions. One of the most important is nuclear shadowing, i.e., the modification of the quark and gluon structure functions for nucleons inside nuclei (see e.g., Refs.~\cite{Eskola:2009uj,Kovarik:2015cma,Eskola:2016oht}). This effect modifies the probability for a quark or a gluon to carry a given fraction $x$ of the momentum of the nucleon. It affects the elementary production cross section for the creation of the c$\overline{\rm c}$ pair that will eventually form a charmonium state. Modifications of the initial state of the nucleus are also addressed by calculations incorporating parton saturation, a coherent effect involving low-$x$ quarks and gluons, described by the Colour Glass Condensate (CGC) effective theory~\cite{Iancu:2003xm}. In addition to these mechanisms, a coherent energy-loss effect involving partons in the initial and final state can also lead to a modification of the parton kinematics and consequently to a change in the quarkonium yields with respect to elementary nucleon-nucleon collisions~\cite{Arleo:2012rs}. Finally, once produced, the charmonium state could be dissociated via inelastic interactions with the surrounding  nucleons~\cite{Gerschel:1988wn}. This process, which plays a dominant role among CNM effects at low collision energy~\cite{Alessandro:2003pi,Adare:2012qf}, should become negligible at the LHC, where the crossing time of the two nuclei is much shorter than the formation time of the resonance~\cite{Hufner:2000jb,Kharzeev:1999bh,Adam:2016ohd}.

The CNM effects introduced above are present in nucleus-nucleus collisions, but can be more directly investigated by studying proton-nucleus collisions, where the contribution of hot-matter effects are thought to be negligible. Previous results from \mbox{p--Pb} collisions at $\sqrt{s_{\rm NN}}=5.02$ TeV from ALICE~\cite{Abelev:2013yxa,Adam:2015iga,Adam:2015jsa}, LHCb~\cite{Aaij:2013zxa} and CMS~\cite{Sirunyan:2017mzd} have shown a significant suppression of the J/$\psi$ yield at forward rapidity (p-going direction) and low to intermediate $p_{\rm T}$ ($\lesssim 5$ GeV/$c$). No  significant effects, or at most a slight enhancement, were seen at high $p_{\rm T}$ and at backward $y_{\rm {cms}}$ (Pb-going direction). The results were compared to theoretical calculations that include various combinations of all the effects mentioned in the previous paragraph, except charmonium dissociation in cold nuclear matter~\cite{Albacete:2013ei,Arleo:2013zua,Ferreiro:2014bia,Ma:2015sia,Ducloue:2015gfa,Lansberg:2016deg}. A good agreement with the models was found, indicating on the one hand that mechanisms like shadowing, CGC-related effects and coherent energy loss can account for the observed nuclear effects, and on the other hand that final state break-up processes in nuclear matter have a negligible influence. It should be noted that the model of Ref.~\cite{Ferreiro:2014bia} includes the effects of the interaction of charmonia with a dense hadronic medium possibly created in \mbox{p--Pb} collisions. However, such a medium may be expected to dissociate the weakly bound $\psi(2S)$ state~\cite{Adam:2016ohd}, but should have little or no effect on the strongly bound J/$\psi$ meson. 

In 2016, \mbox{p--Pb} collisions at $\sqrt{s_{\rm NN}}=8.16$ TeV were delivered by the LHC. The interest in J/$\psi$ studies at this energy is threefold: first, a significantly larger integrated luminosity with respect to studies performed at $\sqrt{s_{\rm NN}}=5.02$ TeV~\cite{Abelev:2013yxa,Adam:2015iga,Adam:2015jsa} has become available in ALICE, allowing a more detailed comparison to model calculations and an extended $p_{\rm T}$ reach. Second, by varying the collision energy, it is possible to extend the investigations of shadowing and other CNM effects to a partly different $x$ range. Finally, studies of various physics observables in \mbox{p--Pb} and high-muliplicity pp collisions at the LHC have shown effects such as long-range two-particle correlations~\cite{Khachatryan:2016txc,CMS:2012qk,Aad:2015gqa,Aad:2012gla,Abelev:2012ola,ABELEV:2013wsa} and an enhancement of strange and multi-strange hadron production~\cite{ALICE:2017jyt}, already seen in \mbox{Pb--Pb} collisions. These effects are usually connected with the formation of an extended system of strongly interacting particles. Concerning the specific case of charmonium production, in addition to the observations discussed above, long-range correlation structures in J/$\psi$ production were recently observed in \mbox{p-Pb} collisions at $\sqrt{s_{\rm NN}}=8.16$ TeV~\cite{Acharya:2017tfn}. Furthermore, for the weakly bound $\psi(2S)$ a suppression signal, on top of the CNM effects discussed in the previous paragraphs, was seen in \mbox{p--Pb} and related to the resonance break-up in the medium created in such collisions~\cite{Abelev:2014zpa,Adam:2016ohd}. As mentioned above, no extra suppression needs to be introduced for the strongly bound J/$\psi$  in order to reproduce the experimental observations at $\sqrt{s_{\rm NN}}=5.02$ TeV. However, higher energy \mbox{p--Pb} collisions may create a more extended and longer-lived medium, which might lead to a suppression effect also on the J/$\psi$.

In this paper, we report ALICE results on cross sections and nuclear modification factors for inclusive J/$\psi$ production in \mbox{p--Pb} collisions at $\sqrt{s_{\rm NN}}=8.16$ TeV, in the rapidity regions $2.03<y_{\rm {cms}}<3.53$ and $-4.46<y_{\rm {cms}}<-2.96$, and for $p_{\rm T}<20$ GeV/$c$. In Sec.~2, the experimental apparatus, the data sample and the event selection criteria are presented. Section 3 contains a description of the analysis procedure, including a discussion of the evaluation of the systematic uncertainties. The results and their comparison to theoretical models, to recent LHCb results~\cite{Aaij:2017cqq} and to $\sqrt{s_{\rm NN}}=5.02$ TeV data are shown in Sec.~4, while conclusions are drawn in Sec.~5. 

\section{Experimental apparatus, data sample and event selection}
\label{section:Apparatus}

The ALICE detector design and performance are extensively described in~\cite{Aamodt:2008zz,Abelev:2014ffa}. 
The analysis presented here is based on the detection of muons in the ALICE forward muon spectrometer~\cite{Aamodt:2011gj}, which includes five tracking stations (Cathode Pad Chamber detectors), followed by two triggering stations (Resistive Plate Chamber detectors). An absorber, 10 interaction-length ($\lambda_{\rm I}$) thick and made of carbon, concrete and steel, positioned in front of the tracking system, filters out most hadrons produced in the collision. A second  (7.2 $\lambda_{\rm I}$ thick) iron absorber, positioned between the tracking and the triggering system, absorbs secondary hadrons escaping the first absorber and low-momentum muons. Finally, a 3 T$\cdot$m dipole magnet, positioned in the region of the third tracking station, provides the track bending for momentum evaluation. Particles are detected in the pseudo-rapidity range $-4<\eta<-2.5$ in the laboratory system and muon triggering is performed with a programmable transverse momentum threshold, set to $p_{\mu,\rm T}=0.5$ GeV/$c$ for the data sample analysed in this paper. The trigger threshold is not sharp, and the single muon trigger efficiency reaches its plateau value ($\sim$ 96\%) at $p_{\mu,\rm T}\sim 1.5$ GeV/$c$. 
 
In addition to the muon spectrometer, four other sets of detectors play an important role for this analysis. The Silicon Pixel Detector (SPD)~\cite{Aamodt:2010aa}, with its two layers covering the pseudo-rapidity intervals $|\eta|<2$ and $|\eta|<1.4$, is part of the ALICE central barrel and is used to reconstruct the primary vertex. A coincidence of a signal in the two V0 scintillator detectors~\cite{Abbas:2013taa}, covering $2.8<\eta< 5.1$ and $-3.7<\eta<-1.7$, provides a minimum-bias (MB) trigger. The luminosity determination is obtained from the V0 information and, independently, using the T0 Cherenkov detectors~\cite{Bondila:2005xy}, which cover $4.6<\eta <4.9$ and $-3.3<\eta<3.0$. Finally, the timing information from the V0 and the Zero Degree Calorimeters (ZDC)~\cite{ALICE:2012aa} is used to remove beam-induced background. 

The trigger condition used in the analysis is a $\mu\mu-{\rm MB}$ trigger formed by the coincidence of the MB trigger and an unlike-sign dimuon trigger. By taking data in two configurations of the beams corresponding to either protons or Pb ions going towards the muon spectrometer, it was possible to cover the dimuon rapidity ranges $2.03<y_{\rm {cms}}<3.53$  and $-4.46<y_{\rm {cms}}<-2.96$, respectively. The two configurations are also referred to as \mbox{p--Pb} and \mbox{Pb--p} in the following. 


The data samples used in this analysis correspond to an integrated luminosity ${\mathcal L}_{\rm int}^{\text {pPb}}= 8.4\pm 0.2\, {\rm nb}^{-1}$ for \mbox{p--Pb},  and ${\mathcal L}_{\rm int}^{\text {Pbp}}= 12.8\pm 0.3\, {\rm nb}^{-1}$ for \mbox{Pb--p} collisions~\cite{ALICE-PUBLIC-2018-002}. These values are larger by about a factor 2 with respect to $\sqrt{s_{\rm NN}}=5.02$ \mbox{p--Pb} collision data~\cite{Abelev:2013yxa}.

The selection criteria used by ALICE in previous J/$\psi$ analyses~\cite{Abelev:2013yxa,Adam:2015iga} have been applied. Namely, both muons belonging to the pair must have $-4<\eta_\mu<-2.5$, to reject tracks at the edges of the acceptance. In addition, each muon must have $17.6< R_{\rm abs} <89.5$ cm, where $R_{\rm abs}$ is the radial transverse position of the muon tracks at the end of the absorber, to remove tracks crossing its thicker region, where energy loss and multiple scattering effects are more important. Finally, 
each track reconstructed in the tracking chambers of the muon spectrometer has to match a trigger track reconstructed in the trigger system.

\section{Data analysis}\label{section:analysis}

The analysis procedure is the same for the two data sets discussed in this paper, and very similar to the one reported in Refs.~\cite{Abelev:2013yxa,Adam:2015iga} for the $\sqrt{s_{\rm NN}}=5.02$ TeV \mbox{p--Pb} sample. The inclusive J/$\psi$ production cross section was obtained from

\begin{equation}
\frac{{\rm d}^{2}\sigma^{\rm J/\psi}_{\rm pPb}}{{\rm d}y_{\rm {cms}}{\rm d}p_{\rm T}} =\frac{N_{\rm J/\psi}(\Delta y_{\rm cms},\Delta p_{\rm T})}
{\mathcal{L}_{\rm int}^{\rm pPb} \cdot (A\times\varepsilon)_{(\Delta y_{\rm cms},\Delta p_{\rm T})} \cdot {\rm B.R.}({\rm J}/\psi\rightarrow \mu{^+}\mu{^-}) \cdot \Delta y_{\rm {cms}} \cdot \Delta p_{\rm T}}
\end{equation}

where $N_{\rm J/\psi}(\Delta y_{\rm cms},\Delta p_{\rm T})$ is the number of reconstructed J/$\psi$ in the $(\Delta y_{\rm cms},\Delta p_{\rm T})$ interval under consideration, $(A\times\varepsilon)_{(\Delta y_{\rm cms},\Delta p_{\rm T})}$ is the corresponding product of acceptance times reconstruction efficiency, ${\rm B.R.}({\rm J}/\psi\rightarrow \mu{^+}\mu{^-})=5.961\pm 0.033$\% is the branching ratio for the decay to a muon pair~\cite{Patrignani:2016xqp} and ${\mathcal{L}}_{\rm int}^{\rm pPb}$ is the integrated luminosity for the data sample under study.

\begin{figure}
\begin{center}
\includegraphics[width=0.48\linewidth]{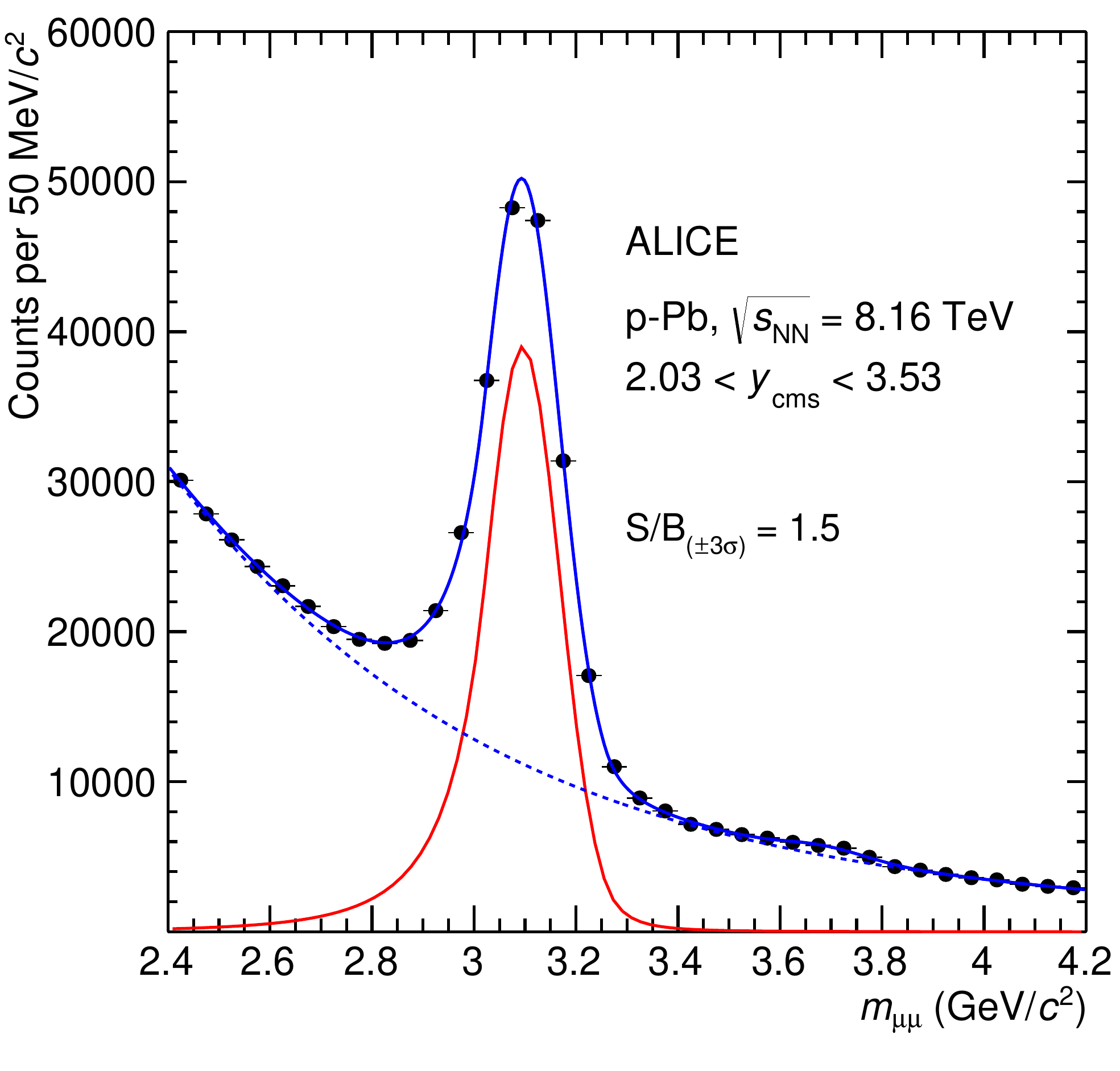}
\includegraphics[width=0.48\linewidth]{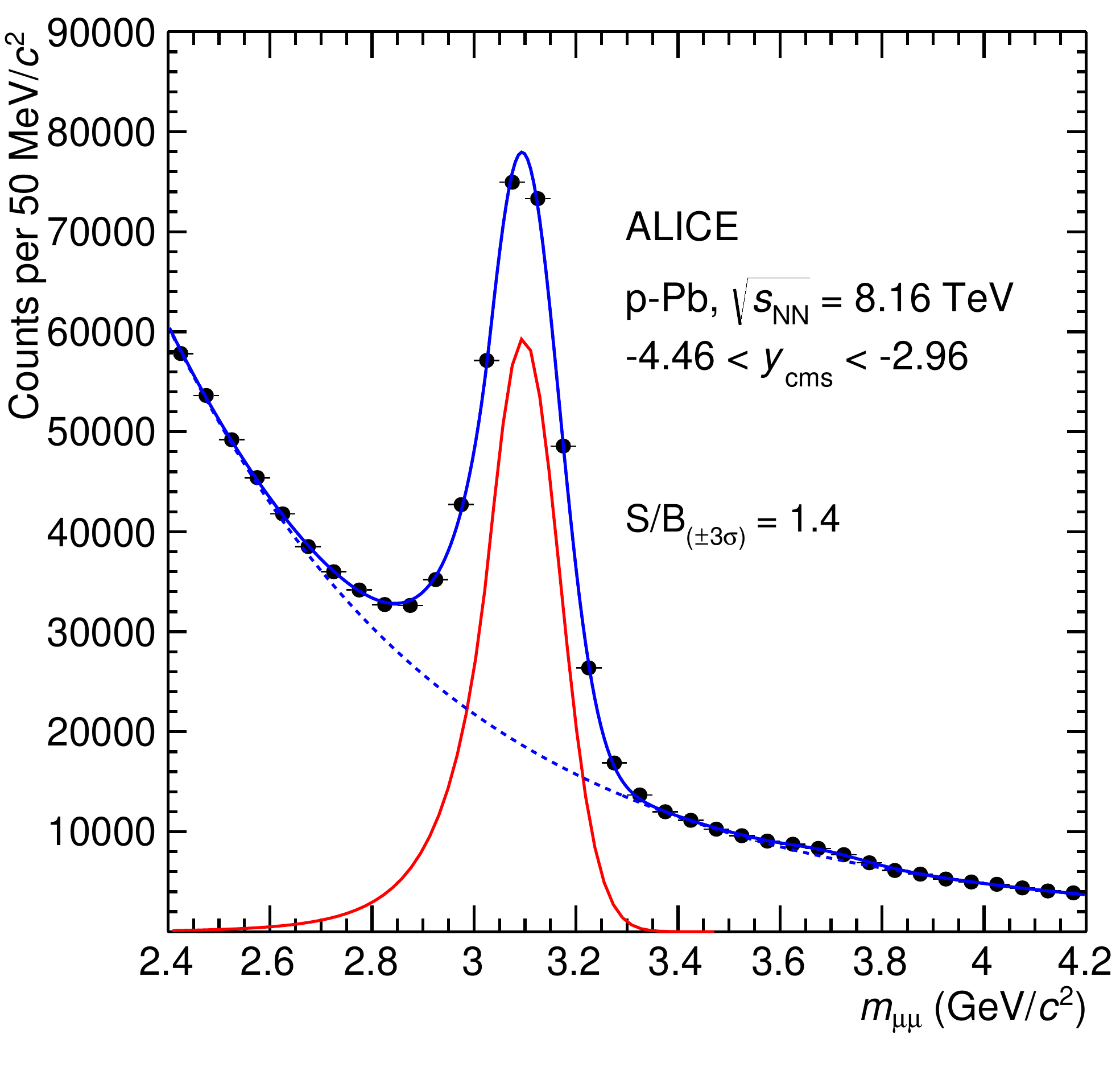}
\caption{Fits to the invariant mass distributions of opposite-sign dimuons with $p_{\rm T}<20$ GeV/$c$. The left plot refers to $2.03<y_{\rm {cms}}< 3.53$, and that on the right to $-4.46<y_{\rm {cms}}<-2.96$. 
The shapes of both the resonances and the background are also shown.}
\label{fig:fitinvmass}
\end{center}
\end{figure}

The quantities $N_{\rm J/\psi}(\Delta y_{\rm cms},\Delta p_{\rm T})$ were obtained through fits to the invariant mass spectra of the opposite-sign muon pairs. The fitting functions are the sum of two resonance contributions (J/$\psi$ and $\psi(2S)$) and a continuum background. For the resonances~\cite{ALICE-PUBLIC-2015-006}, an ``extended'' Crystal Ball (CB2) function was adopted, which accommodates a non-Gaussian tail both on the right and on the left side of the resonance peak.  Alternatively, a pseudo-Gaussian function was used, corresponding to
a resonance Gaussian core around the J/$\psi$ pole and tails on the right and left side of it, parameterised by varying the width of the Gaussian as a function of the mass. The background was described by empirical functions, either with a Gaussian with a mass-dependent width or with an exponential function times a fourth-order polynomial~\cite{ALICE-PUBLIC-2015-006}.
Fits were performed using all the combinations of the signal and background functions, and varying the fitting ranges ($2.2<m_{\mu\mu}<4.5$ GeV/$c^2$ or $2<m_{\mu\mu}<5$ GeV/$c^2$). Figure~\ref{fig:fitinvmass} shows an example of fits to the invariant mass distributions of the \mbox{p--Pb} and \mbox{Pb--p} data samples, for opposite-sign dimuons in the region $p_{\rm T}<20$ GeV/$c$.

When fitting the mass spectra, the  value of the J/$\psi$ mass and its width ($\sigma$) at the pole position are free parameters of the fit. 
The contribution of the $\psi(2S)$ was found to have a negligible impact on the evaluation of $N_{\rm J/\psi}$. 

A study of the influence of the non-Gaussian tails of the shapes of the reconstructed resonance spectra was also performed. The corresponding fit parameters were extracted either from the MC or directly from data. In the latter case, the tail parameters were evaluated either by leaving them as free parameters in the fit to the \mbox{p--Pb} and \mbox{Pb--p} samples, or using values obtained from the corresponding pp data samples at $\sqrt{s}=8$ TeV~\cite{Adam:2015rta} (about the same energy of the collisions under study) or $\sqrt{s}=13$ TeV~\cite{Acharya:2017hjh} (largest data sample collected by ALICE).

The $N_{\rm J/\psi}$ values were finally obtained as the average of the results of all the fits performed. The statistical uncertainties were obtained as the average of the statistical uncertainties over the various fits, while the standard deviations of the $N_{\rm J/\psi}$ distributions were taken as the systematic uncertainties. Typical values of the signal over background ratio in a 3$\sigma$ window around the J/$\psi$ peak range from 1.4 (0.7) to 2.8 (1.4) from low to high $p_{\rm T}$ in \mbox{p--Pb} (\mbox{Pb--p}) collisions.
For the $p_{\rm T}$-integrated data samples ($p_{\rm T}<20$ GeV/$c$), $N_{\rm J/\psi}$ amounts to $(1.67\pm 0.01\pm 0.05)\cdot 10^{5}$ and $(2.52\pm 0.01\pm 0.08)\cdot 10^{5}$ for \mbox{p--Pb} and \mbox{Pb--p} respectively, where the first uncertainty is statistical and the second systematic. The latter quantity (which amounts in percentage terms to $\sim 3$\%) is dominated by the choice of the J/$\psi$ tail parameters.
When extracting $N_{\rm J/\psi}$ in narrower $p_{\rm T}$ and $y_{\rm {cms}}$ ranges, the systematic uncertainties turn out to be similar (from 3\% up to 4\% in the highest $p_{\rm T}$ bins).

The quantity $(A\times\varepsilon)_{(\Delta y_{\rm cms},\Delta p_{\rm T})}$ was evaluated by means of MC simulations, performed separately for each data taking run, in order to follow the evolution of the detector conditions. The  input $p_{\rm T}$ and $y_{\rm {cms}}$ distributions for the J/$\psi$ were tuned directly to the data by means of an iterative procedure. In detail, a first set of differential distributions, corresponding to the results of the measurements performed at $\sqrt{s_{\rm NN}}=5.02 $ TeV~\cite{Adam:2015iga}, was taken as an input to the calculation, and the resulting $(A\times\varepsilon)_{(\Delta y_{\rm cms},\Delta p_{\rm T})}$ values were then used to correct the raw J/$\psi$ distributions obtained from the fits of the invariant mass spectra. The corrected differential distributions were then used as an input for another $(A\times\varepsilon)_{(\Delta y_{\rm cms},\Delta p_{\rm T})}$ calculation, and so on. Convergence was reached at the second iteration. The $p_{\rm T}$-integrated values of $(A\times\varepsilon)$ are $0.2646\pm 0.0001$ (\mbox{p--Pb}) and $0.2349\pm 0.0001$ (\mbox{Pb--p}), where the quoted uncertainties are statistical. 

The systematic uncertainties are related to the corresponding uncertainties on the trigger and tracking efficiencies, as well as to the choice of the input distributions. Concerning the efficiencies, for the muon trigger the procedure already used for the analysis of \mbox{p--Pb} data at $\sqrt{s_{\rm NN}}=5.02$ TeV was followed~\cite{Abelev:2013yxa,Adam:2015iga}. The response function of the muon trigger obtained in MC and in data was used for the J/$\psi$  $(A\times\varepsilon)_{(\Delta y_{\rm cms},\Delta p_{\rm T})}$ calculation. Integrating over $p_{\rm T}$, a difference of 2.4\% (2.9\%) on the trigger efficiency for J/$\psi$ was estimated in \mbox{p--Pb} (\mbox{Pb--p}) collisions. The difference  can become as high as 4\% for low-$p_{\rm T}$ J/$\psi$. A 1\% contribution due to the uncertainty on the intrinsic efficiency of the muon-trigger detectors was then added in quadrature to the quoted uncertainties. For the tracking efficiency, the corresponding systematic uncertainty was calculated by comparing the efficiencies evaluated in data and MC. The efficiency of each tracking plane was obtained using the redundancy of the tracking system (two independent planes per station). Then, the single muon 
tracking efficiencies were calculated according to the tracking algorithm, and finally combined, in order to get the dimuon tracking efficiency. The estimated value of the systematic uncertainty on the tracking efficiency is 1\% (2\%) for $p_{\rm T}$-integrated J/$\psi$ production in \mbox{p--Pb} (\mbox{Pb--p}) collisions, and shows no appreciable dependence on the dimuon kinematics. A further systematic uncertainty, related to the choice of the $\chi^2$ cut applied to the matching of tracks reconstructed in the muon tracking and triggering systems, was also included. Its value is 1\%, independent of $p_{\rm T}$ and $y_{\rm {cms}}$.
Finally, the choice of the MC input distributions was found to induce a 0.5\% systematic uncertainty on the acceptance calculation for the $p_{\rm T}$-integrated data samples. This effect is due to the statistical uncertainty on the measured $y_{\rm {cms}}$ and $p_{\rm T}$ distributions that were used for the calculation, and to possible correlations between the distributions in the two kinematic variables. The maximum value of this uncertainty becomes 3\% at very low $p_{\rm T}$ ($<$1~GeV/$c$).

The integrated luminosities for the two data samples were obtained from   $\mathcal{L}_{\rm int} = N_{\rm MB}/\sigma_{\rm MB}$ where $N_{\rm MB}$ is the number of MB events and $\sigma_{\rm MB}$ the cross section corresponding to the MB trigger condition. The latter quantity was evaluated from a van der Meer scan, obtaining $2.09\pm 0.03$~b for \mbox{p--Pb} and $2.10 \pm 0.04$~b for \mbox{Pb--p}~\cite{ALICE-PUBLIC-2018-002}. The $N_{\rm MB}$ quantity was estimated as $N_{\mu\mu-{\rm MB}}\cdot F_{\rm norm}$, where $N_{\mu\mu-{\rm MB}}$ is the number of analysed dimuon triggers and $F_{\rm norm}$ is the inverse of the probability of having a triggered dimuon in a MB event. $F_{\rm norm}$ was calculated using the event trigger information, as the ratio between the number of collected MB triggers and the 
number of times the MB condition is verified together with the dimuon trigger condition, with the latter information obtained from the level-0 trigger mask. 
The $F_{\rm norm}$ values were evaluated, and corrected for the small pile-up contribution to the MB sample ($\sim 3$\% on average), for each run and finally averaged using as a weight the number of $\mu\mu-{\rm MB}$ triggers. In this way one obtains $F^{\rm pPb}_{\rm norm}=679\pm 7$ and $F^{\rm Pbp}_{\rm norm}=371\pm 4$. The quoted uncertainties (1\%) are systematic and were obtained by comparing the results of the evaluation described above with an alternative method based on the information of the trigger scalers~\cite{Abelev:2013yxa}. Statistical uncertainties on $F_{\rm norm}$ are negligible.
 
The nuclear effects on J/$\psi$ production in \mbox{p--Pb} collisions were estimated via the nuclear modification factor, defined as:

\begin{equation}
R_{\rm pPb}(y_{\rm cms},p_{\rm T})=\frac{{\rm d}^{2}\sigma^{\rm J/\psi}_{\rm pPb}/{\rm d}y_{\rm cms}{\rm d}p_{\rm T}}{A_{\rm Pb} \cdot
{\rm d}^{2}\sigma^{\rm J/\psi}_{\rm pp}/{\rm d}y_{\rm cms}{\rm d}p_{\rm T}}
\end{equation}

where the \mbox{p--Pb} production cross section is normalised to the corresponding quantity for pp collisions times the atomic mass number of the Pb nucleus ($A_{\rm Pb}=208$).

The reference pp cross section was evaluated starting from the available results for forward-$y_{\rm {cms}}$ inclusive J/$\psi$ production at $\sqrt{s}=8$ TeV from ALICE~\cite{Adam:2015rta} and LHCb~\cite{Aaij:2013yaa}. These results are in fair agreement, as their maximum difference is 1.4$\sigma$, in the region close to $y_{\rm {cms}}=2.5$. Since the ALICE pp data cover a different $y_{\rm {cms}}$-range ($2.5<y_{\rm {cms}}<4$) with respect to those accessible in \mbox{p--Pb} and \mbox{Pb--p} collisions, a rapidity extrapolation by $\sim \pm0.5$ $y$-units was performed to match the kinematic window of the various samples, following the procedure described in~\cite{LHCb-CONF-2013-013}. In addition, a $\sqrt{s}$-interpolation~\cite{Acharya:2017hjh} was performed to account for the small difference in the centre-of-mass energy between pp and proton-nucleus collisions. 

The rapidity extrapolation was performed on the ALICE data using three different functions (Gaussian, 2$^{\rm nd}$ and 4$^{\rm th}$ degree polynomials) and taking the weighted average of the extrapolated values. The associated systematic uncertainty was calculated as the maximum difference between the results obtained with the different functions. Typical values are $\sim 2-3$\%, reaching a maximum of $\sim 25\%$ at the very edge of the extrapolation region. For LHCb, the same procedure was used in order to match the rapidity binning of the \mbox{p--Pb} and \mbox{Pb--p} data. The procedure corresponds in this case to an interpolation, because of the larger rapidity acceptance ($2<y_{\rm {cms}}<4.5$) of LHCb. Finally, the weighted average of the ALICE/LHCb based extrapolations/interpolations was calculated, and a small correction factor ($1.5$\%), obtained via a $\sqrt{s}$-interpolation of data at various centre-of-mass energies, was introduced to account for the slight centre-of-mass energy difference between \mbox{p--Pb} ($\sqrt{s_{\rm NN}}=8.16$ TeV) and \mbox{pp} data ($\sqrt{s}=8$ TeV).

For the $p_{\rm T}$-differential studies, the reference pp cross section was  obtained as a weighted average of the ALICE and LHCb $p_{\rm T}$-differential cross sections at $\sqrt{s}=8$ TeV~\cite{Adam:2015rta,Aaij:2013yaa}, extrapolated/interpolated to the proton-nucleus rapidity domains. The ALICE  values, which are extrapolated beyond the measured pp rapidity range, were also corrected by $p_{\rm T}$-dependent factors, which account for the softening/hardening of the $p_{\rm T}$-differential cross section when $y_{\rm {cms}}$ increases/decreases, and were calculated from the LHCb pp results on the $p_{\rm T}$-differential inclusive J/$\psi$ cross section in narrow $y_{\rm {cms}}$ bins~\cite{Aaij:2013yaa}. Since the $p_{\rm T}$ coverage of pp data at $\sqrt{s}=8$ TeV by LHCb extends only up to $p_{\rm T}=14$ GeV/$c$, a linear extrapolation of the correction factors up to $p_{\rm T}=20$ GeV/$c$ was performed. The size of this correction is $<10\%$ for $p_{\rm T}\lesssim 6$ GeV/$c$ and increases up to $\sim 40$\% in the highest $p_{\rm T}$ bin. The uncertainty associated with this correction factor is small ($1-2$\%), thanks to the very good accuracy of the LHCb results. Finally, the effect of the slight centre-of-mass energy difference between proton-nucleus and \mbox{pp} data sets  ranges from 1\% to 3.5\% when increasing $p_{\rm T}$. 

Table 1 summarises the systematic uncertainties on the various contributions  entering the cross section and the nuclear modification factor determination. The uncertainty on the integrated luminosity is the sum in quadrature of the uncertainties on $\sigma_{\rm MB}$~\cite{ALICE-PUBLIC-2018-002} and $F_{\rm norm}$. 
The fractions correlated/uncorrelated between \mbox{p--Pb} and \mbox{Pb--p} measurements are separately quoted. 
\begin{table}
\begin{center}
\begin{tabular}{c|c|c|c|c|c|c}
\hline
& \multicolumn{3}{c|}{p--Pb ($2.03<y_{\rm {cms}}<3.53$)} & \multicolumn{3}{c}{Pb--p ($-4.46<y_{\rm {cms}}<-2.96$)} \\
\hline
Source & Integrated & vs $p_{\rm T}$ & vs $y_{\rm {cms}}$ & Integrated & vs $p_{\rm T}$ & vs $y_{\rm {cms}}$\\
\hline
\hline
Signal extraction & 3.1\% & 2.9--4.2\% & 3.1--3.2\% & 3.4\% & 2.7--4.0\% & 3.1--3.3\% \\
\hline
MC input & 0.5\% & 1--3\% & 1\% & 0.5\% & 1--2\% & 1--2\%\\
\hline
Tracking efficiency & 1\% & 1\% & 1\% & 2\% & 2\% & 2\%\\
\hline
Trigger efficiency & 2.6\% & 1.4--4.1\% & 2.2--4.1\% & 3.1\% & 1.4--4.1\% & 3.2--4.1\%\\
\hline
Matching efficiency & 1\% & 1\% & 1\% & 1\% & 1\% & 1\%\\
\hline
$\mathcal{L}_{\rm int}^{\rm pPb}$ (uncorrelated) & \multicolumn{3}{c|}{2.1\%} & \multicolumn{3}{c}{2.2\%} \\
\hline
$\mathcal{L}_{\rm int}^{\rm pPb}$ (correlated) & \multicolumn{3}{c|}{0.5\%} & \multicolumn{3}{c}{0.7\%}\\
\hline
B.R.(${\rm J}/\psi\rightarrow \mu^+\mu^-$) & \multicolumn{6}{c}{0.6\%} \\
\hline
pp reference (unc.) & 1.5\% & 3.5--17.0\% & 1.6--3.5\% & 1.8\% & 3.6--15.4\% & 1.8--5.9\% \\
\hline
pp reference (corr.) & \multicolumn{6}{c}{7.1\%}\\
\hline

\end{tabular}
\caption{Summary of systematic uncertainties on the calculation of cross sections and nuclear modification factors. Uncertainties on signal extraction, MC input and efficiencies are considered as uncorrelated  over $p_{\rm T}$ and $y_{\rm cms}$.
The uncertainties on the luminosity and on the pp reference result from the combination of two contributions, one uncorrelated and the other correlated, which are separately quoted. The uncorrelated uncertainty on luminosity includes the contribution of the systematic uncertainty on $F_{\rm norm}$ as well as a 1.1\% (0.6\%) contribution due to the difference between the luminosities obtained with the V0 and T0 detectors.}
\end{center}
\label{tab:1}
\end{table}

\section{Results}

\begin{figure}
\begin{center}
\includegraphics[width=0.7\linewidth]{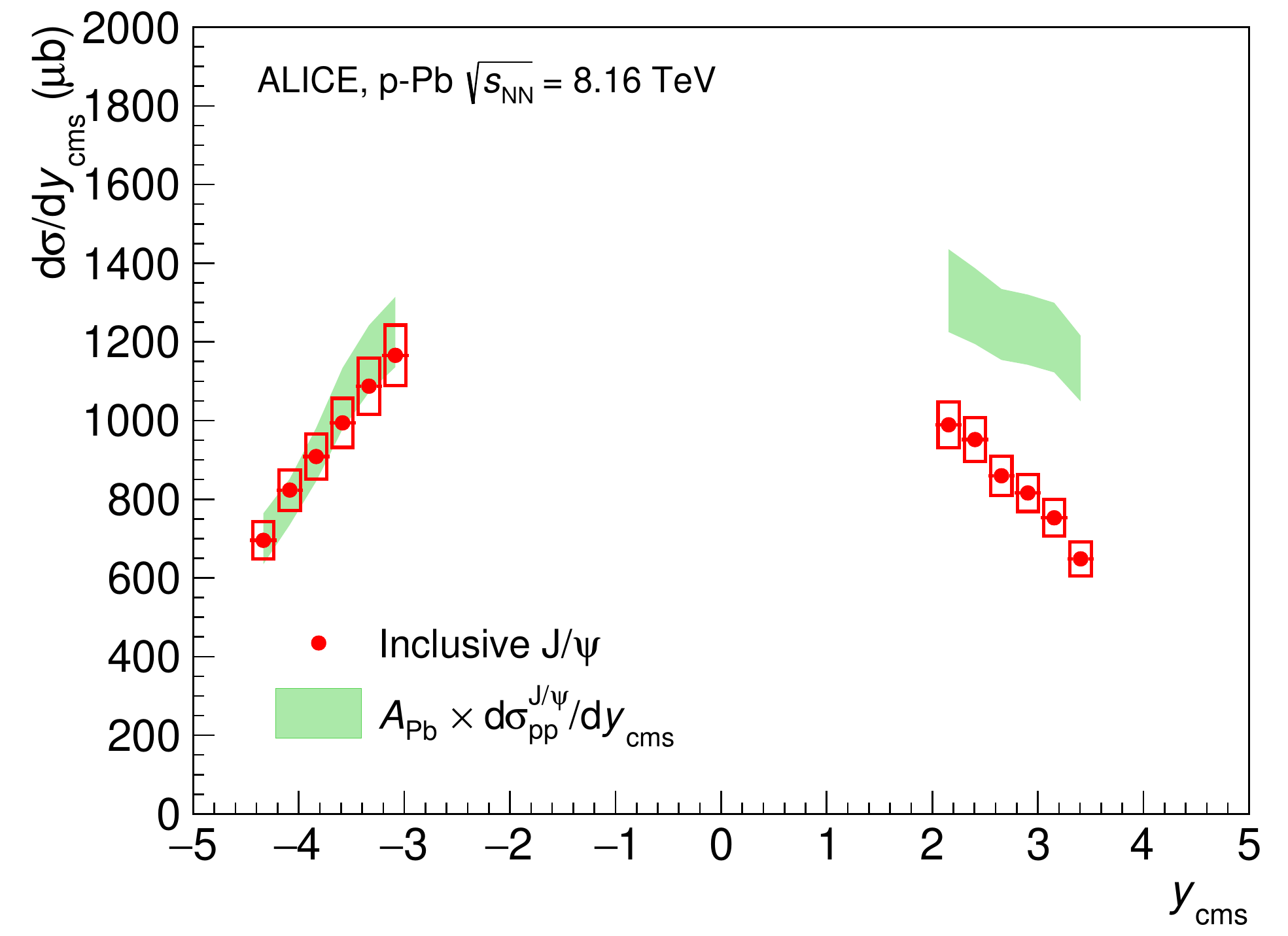}
\caption{The $y$-differential inclusive J/$\psi$ production cross section in \mbox{p--Pb} and \mbox{Pb--p} collisions at $\sqrt{s_{\rm NN}}=8.16$ TeV. The vertical error bars (not visible because smaller than the symbols) represent the statistical uncertainties, the boxes around the points the systematic uncertainties. The horizontal bars correspond to the bin size. The values of the reference pp cross sections, obtained through the interpolation/extrapolation procedure described in Sec.~3, scaled by $A_{\rm Pb}$, are shown as bands.}
\label{fig:sigmavsy}
\end{center}
\end{figure}

\begin{figure}
\begin{center}
\includegraphics[width=0.7\linewidth]{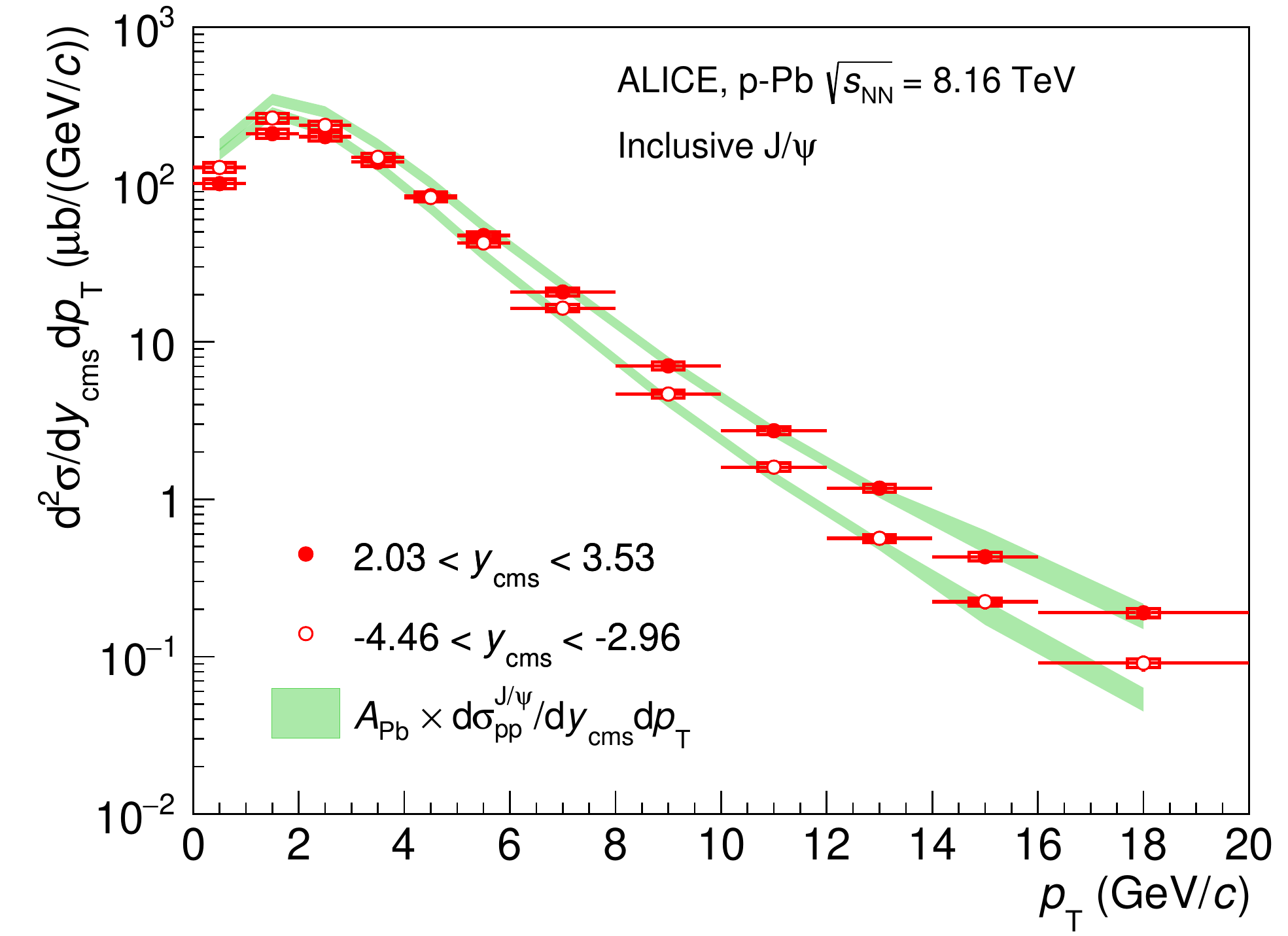}
\caption{The $p_{\rm T}$-differential inclusive J/$\psi$ production cross section in \mbox{p--Pb} and \mbox{Pb--p} collisions at $\sqrt{s_{\rm NN}}=8.16$ TeV. The vertical error bars (not visible because smaller than the symbols) represent the statistical uncertainties, the boxes around the points the systematic uncertainties. The horizontal bars correspond to the bin size. The  values of the reference pp cross sections, obtained through the interpolation/extrapolation procedure described in Sec.~3, scaled by $A_{\rm Pb}$, are shown as bands.}
\label{fig:sigmavspt}
\end{center}
\end{figure}

In Fig.~\ref{fig:sigmavsy} the differential cross sections are presented for inclusive J/$\psi$ production as a function of rapidity in \mbox{p--Pb} and \mbox{Pb--p} collisions, integrated over the transverse momentum interval $p_{\rm T}<20$ GeV/$c$. The same figure shows the reference cross sections for pp collisions, obtained through the interpolation procedure described in Sec.~3 and scaled by $A_{\rm Pb}$.
Figure~\ref{fig:sigmavspt} reports the \mbox{p--Pb} differential cross sections as a function of $p_{\rm T}$, separately for the forward ($2.03<y_{\rm {cms}}<3.53$) and backward ($-4.46<y_{\rm {cms}}<-2.96$) rapidity regions, where  the corresponding pp cross sections scaled by $A_{\rm Pb}$ are also superimposed.
The comparison of proton-nucleus and scaled pp cross sections shows that at forward $y_{\rm cms}$ a suppression of the inclusive J/$\psi$ production
is visible, while no significant nuclear effects can be seen at backward $y_{\rm {cms}}$.

\begin{figure}
\begin{center}
\includegraphics[width=0.7\linewidth]{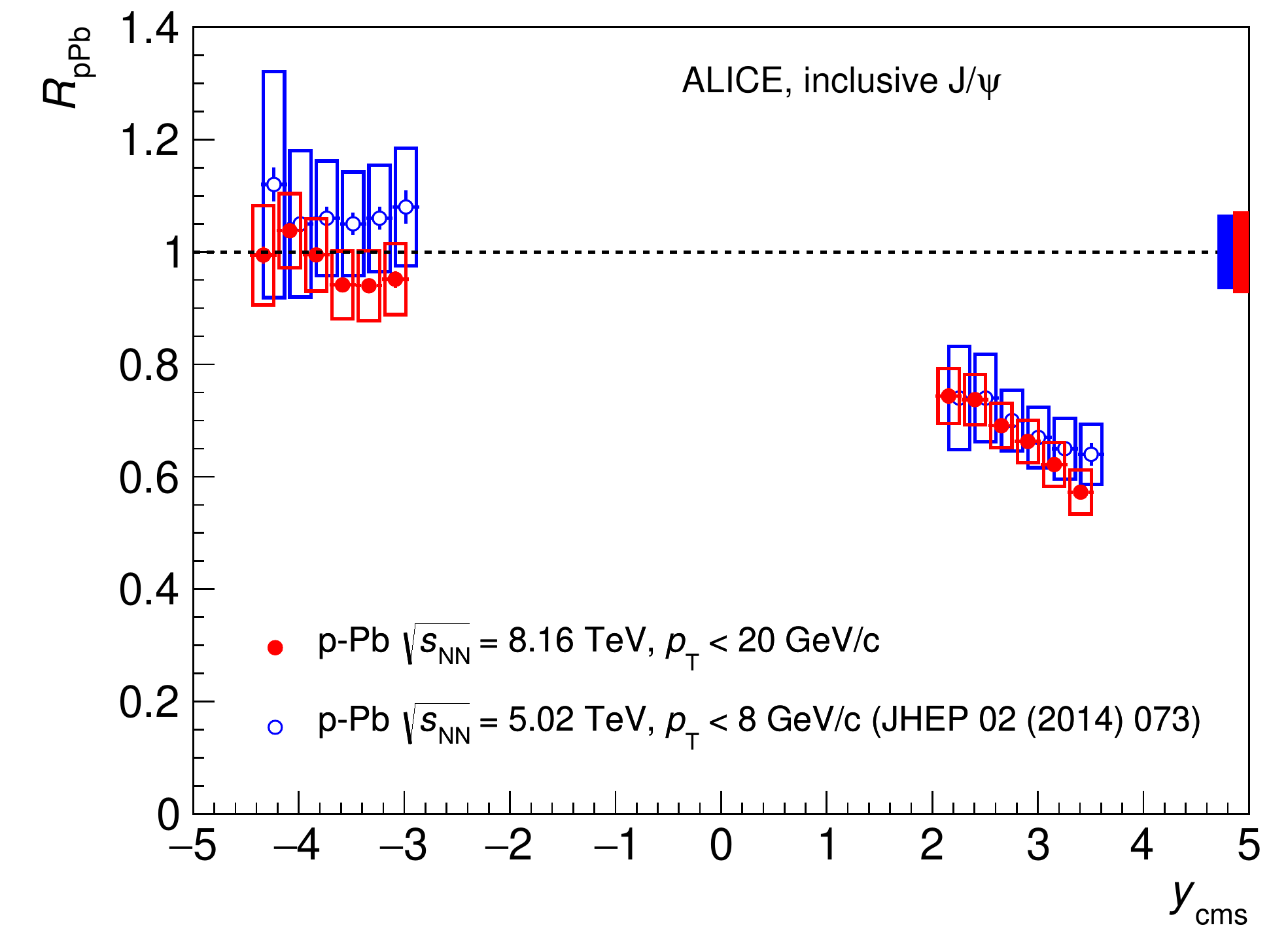}
\caption{The inclusive J/$\psi$ nuclear modification factor in \mbox{p--Pb} and \mbox{Pb--p} collisions at $\sqrt{s_{\rm NN}}=8.16$ TeV, as a function of $y_{\rm {cms}}$. The horizontal bars correspond to the bin size. The vertical error bars represent the statistical uncertainties, the boxes around the points the uncorrelated systematic uncertainties. Correlated uncertainties are shown as a filled box around unity for each energy. The results are compared with those for \mbox{p--Pb} and \mbox{Pb--p} collisions at $\sqrt{s_{\rm NN}}=5.02$ TeV~\cite{Adam:2015iga}. The latter have been plotted at slightly shifted $y_{\rm cms}$ values, for better visibility.}
\label{fig:rppbvsy}
\end{center}
\end{figure}

\begin{figure}
\begin{center}
\includegraphics[width=0.48\linewidth]{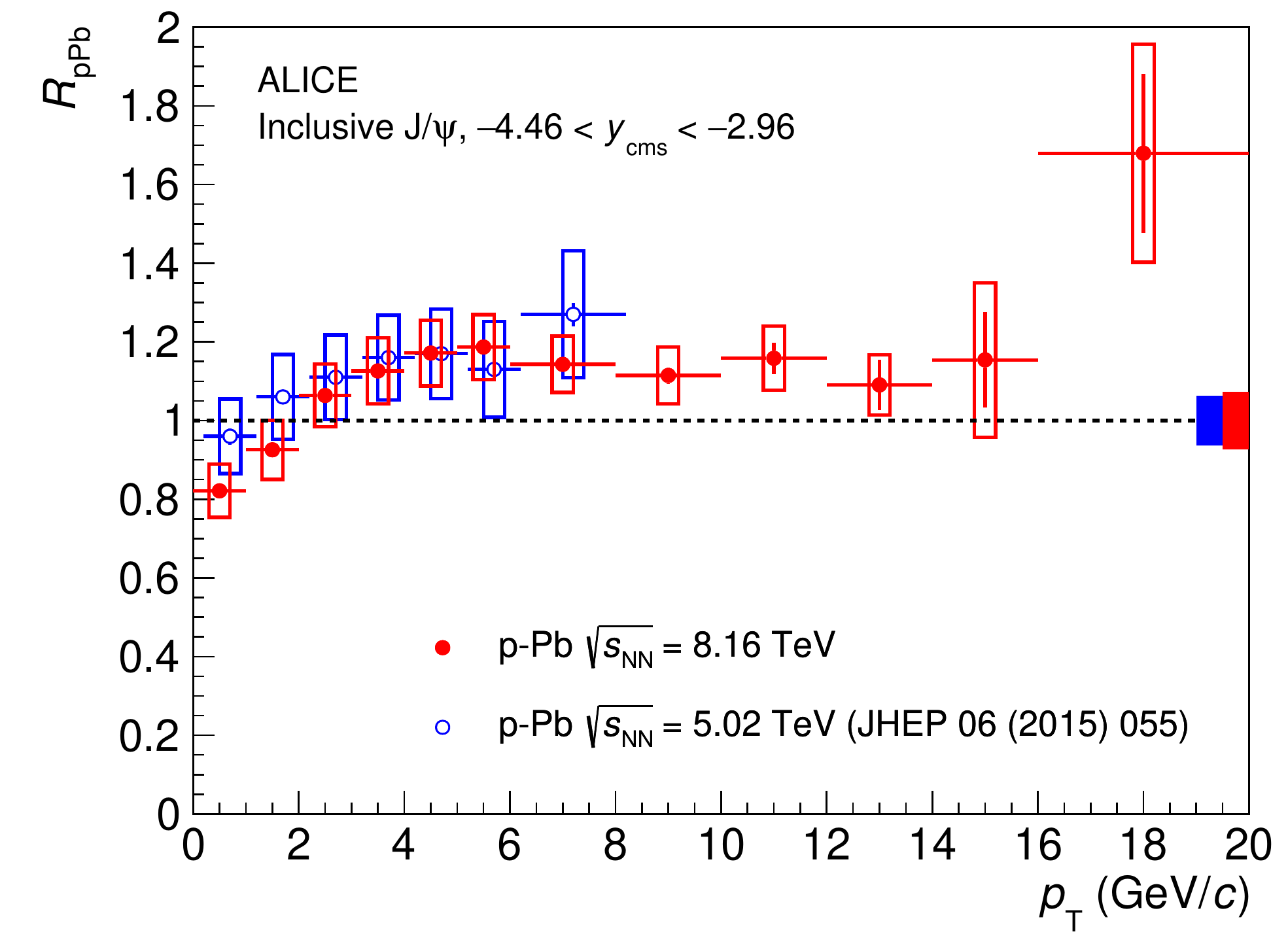}
\includegraphics[width=0.48\linewidth]{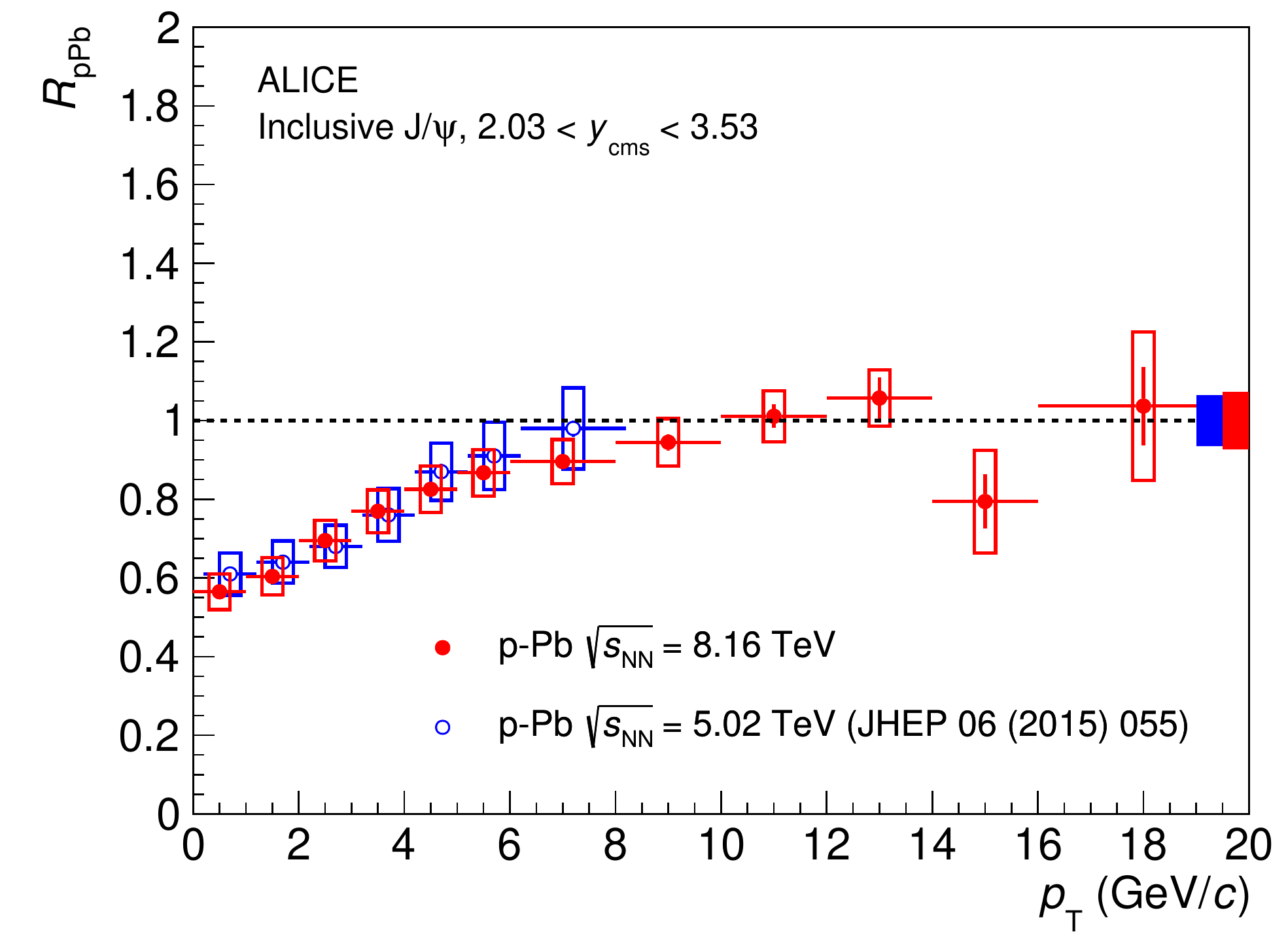}
\caption{The inclusive J/$\psi$ nuclear modification factor in \mbox{Pb--p} (left) and \mbox{p--Pb} (right) collisions at $\sqrt{s_{\rm NN}}=8.16$ TeV, as a function of $p_{\rm T}$. The horizontal bars correspond to the bin size. The vertical error bars represent the statistical uncertainties, the boxes around the points the uncorrelated systematic uncertainties. Correlated uncertainties are shown as a filled box around unity for each energy. The results are compared with those for \mbox{p--Pb} and \mbox{Pb--p} collisions at $\sqrt{s_{\rm NN}}=5.02$ TeV~\cite{Adam:2015iga}. The latter have been plotted at a slightly shifted $p_{\rm T}$, for better visibility.}
\label{fig:rppbvspt}
\end{center}
\end{figure}

Nuclear effects, already visible from the different behaviour of \mbox{p--Pb} and pp-scaled cross sections, are quantified through the nuclear modification factors, shown as a function of $y_{\rm {cms}}$ in Fig.~\ref{fig:rppbvsy} and of $p_{\rm T}$ in Fig.~\ref{fig:rppbvspt}. The results are compared with the corresponding  nuclear modification factors at $\sqrt{s_{\rm NN}}=5.02$ TeV~\cite{Adam:2015iga}. Although the $\sqrt{s_{\rm NN}}=8.16$ TeV data are systematically lower, the difference is not significant given the uncertainties of the measurements. As a function of $y_{\rm {cms}}$, $R_{\rm pPb}$ decreases when moving from the Pb-going to the p-going direction, showing a significant suppression at forward rapidity, while the negative rapidity measurements do not show any significant deviation from unity. As a function of $p_{\rm T}$, an increase is seen at forward $y_{\rm {cms}}$ and the data become compatible with unity for $p_{\rm T}\gtrsim 5$ GeV/$c$. At negative $y_{\rm {cms}}$ an increasing trend is also likely to be present at low transverse momentum, as shown by a fit in the region $p_{\rm T}<4$ GeV/$c$ with a constant function, which gives $\chi^2/{\rm ndf}=3.3$. For $p_{\rm T}>4$ GeV/$c$ the nuclear modification factor is systematically larger than 1, but compatible with unity within 1.9$\sigma$. 

Concerning the compatibility of the results at the two energies, the integration over different $p_{\rm T}$ ranges ($p_{\rm T}<8$ GeV/$c$ for $\sqrt{s_{\rm NN}}=5.02$ TeV data, $p_{\rm T}<20$ GeV/$c$ at $\sqrt{s_{\rm NN}}=8.16$ TeV) in Fig.~\ref{fig:rppbvsy} leads to only a small relative effect on the nuclear modification factors. In fact, when restricting the integration domain of the $\sqrt{s_{\rm NN}}=8.16$ TeV data to $p_{\rm T}<8$ GeV/$c$ the $R_{\rm pPb}$ values decrease by less than 1.5\%.

The nuclear modification factors integrated over rapidity, separately in the forward and backward regions, are

\begin{eqnarray}
R_{\rm{pPb}}(2.03<y_{\rm {cms}}<3.53) = 0.700 \pm 0.005 {\rm(stat.)} \pm 0.065 {\rm(syst.)}\\
R_{\rm{Pbp}}(-4.46<y_{\rm {cms}}<-2.96) = 1.018 \pm 0.004 {\rm(stat.)} \pm 0.098 {\rm(syst.)} 
\end{eqnarray}

demonstrating that the suppression of the J/$\psi$ production at forward rapidity in \mbox{p--Pb} collisions is a 4.6$\sigma$ effect. The corresponding significance for the $\sqrt{s_{\rm NN}}=5.02$ TeV~\cite{Adam:2015iga} data was 3.9$\sigma$. The ratios of the nuclear modification factors obtained at $\sqrt{s_{\rm NN}}=8.16$ and 5.02 TeV, in the region $p_{\rm T}<8$ GeV/$c$, are

\begin{eqnarray}
R_{\rm{pPb}}(8.16\,{\rm TeV})/R_{\rm{pPb}}(5.02\,{\rm TeV})(2.03<y_{\rm {cms}}<3.53) = 0.987 \pm 0.015 {\rm(stat.)} \pm 0.141 {\rm(syst.)}\\
R_{\rm{Pbp}}(8.16\,{\rm TeV})/R_{\rm{Pbp}}(5.02\,{\rm TeV})(-4.46<y_{\rm {cms}}<-2.96) = 0.938 \pm 0.009 {\rm(stat.)} \pm 0.139 {\rm(syst.)} 
\end{eqnarray}

Both values are compatible with unity. The choice of the $p_{\rm T}$ range for the calculation of the ratios is related to the maximum reach of the $\sqrt{s_{\rm NN}}=5.02$ TeV results. 

\begin{figure}
\begin{center}
\includegraphics[width=0.7\linewidth]{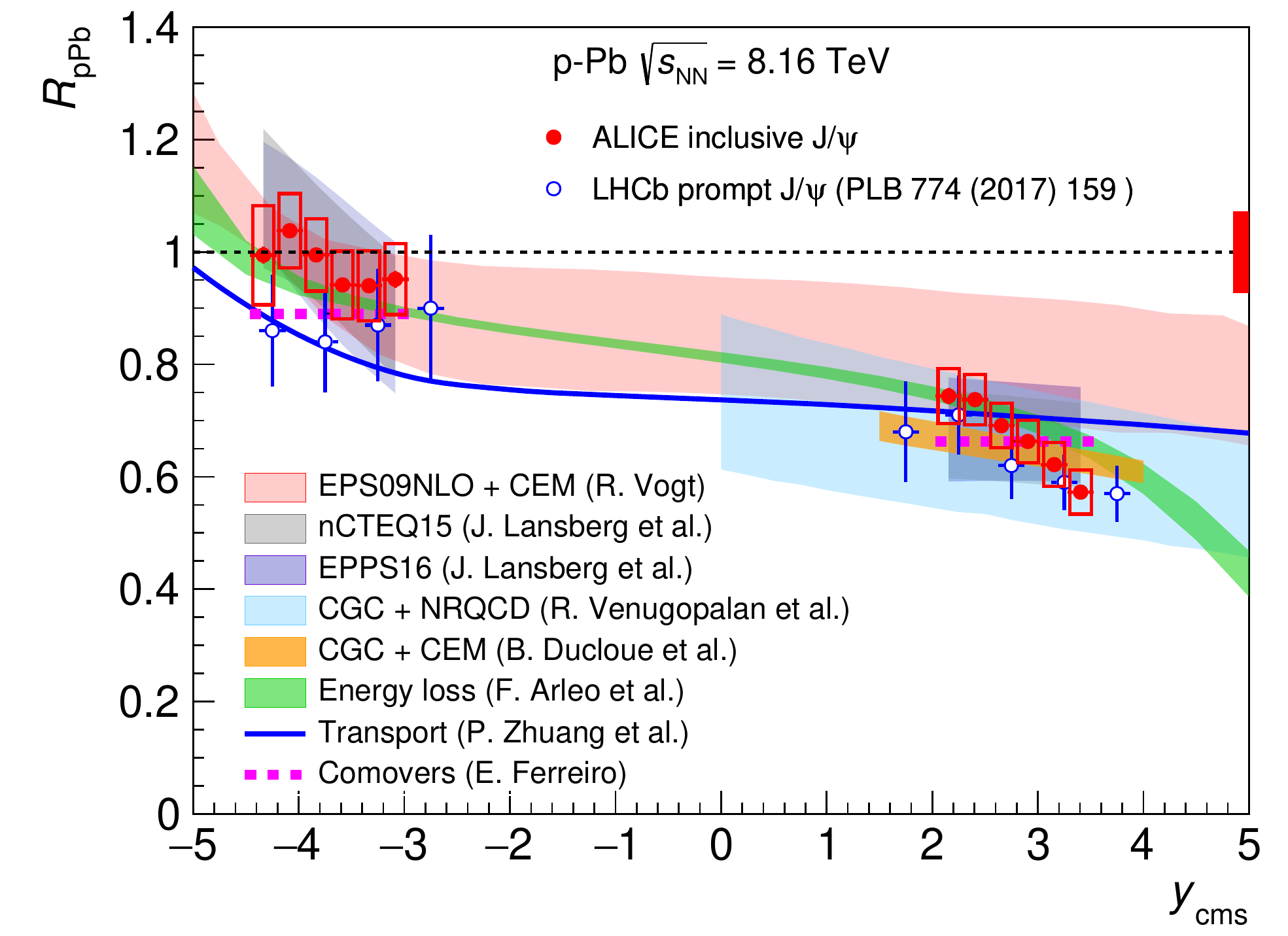}
\caption{Comparison of the ALICE and LHCb~\cite{Aaij:2017cqq} results on the $y_{\rm cms}$-dependence of the J/$\psi$ nuclear modification factors in \mbox{p--Pb} and \mbox{Pb--p} collisions at $\sqrt{s_{\rm NN}}=8.16$ TeV. The horizontal bars correspond to the bin size. For ALICE, the vertical error bars represent the statistical uncertainties, the boxes around the points the uncorrelated systematic uncertainties, and the filled box around unity the correlated uncertainties. For LHCb, the vertical error bars represent the combination of statistical and systematic uncertainties. The results are also compared to  several model calculations~\cite{Albacete:2017qng,Lansberg:2016deg,Ma:2017rsu,Ducloue:2016pqr,Arleo:2014oha,Ferreiro:2014bia,Chen:2016dke} (see text for details).}
\label{fig:rppbvsymodels}
\end{center}
\end{figure}

In Fig.~\ref{fig:rppbvsymodels} the ALICE results are compared to the corresponding LHCb values~\cite{Aaij:2017cqq}, which cover a slightly wider $y_{\rm {cms}}$ range and are integrated up to $p_{\rm T}=14$ GeV/$c$, showing a good agreement between the two measurements. 
The LHCb results refer to prompt J/$\psi$ production, i.e., include decays of higher-mass charmonium states but do not include the contribution from b-hadron decays (non-prompt production). For the region $p_{\rm T}\le 5$ GeV/$c$, which dominates the $p_{\rm T}$-integrated results, the size of the latter contribution amounts to 10--15\% of the inclusive production. 
An estimate of the difference between prompt and inclusive nuclear modification factors, based on LHCb results~\cite{Aaij:2017cqq}, gives a $3-4$\% ($1-2$\%) effect at positive (negative) $y_{\rm {cms}}$.

In Fig.~\ref{fig:rppbvsymodels} a comparison with the results of several theoretical models for prompt J/$\psi$ production is also presented. The results of two calculations based on a pure shadowing scenario (Vogt~\cite{Albacete:2017qng}, Lansberg et al.~\cite{Lansberg:2016deg,Kusina:2017gkz}) show good agreement with data when the nCTEQ15~\cite{Kovarik:2015cma} or EPPS16~\cite{Eskola:2016oht} set of nuclear parton distribution functions (nPDF) is adopted, while using the EPS09~\cite{Eskola:2009uj} set of nPDF leads to a slightly worse agreement at forward $y_{\rm {cms}}$. Calculations based on a CGC approach coupled with various elementary production models are able to reproduce the data in their domain of validity, corresponding to the forward-$y_{\rm {cms}}$ region (Venugopalan et al.~\cite{Ma:2017rsu}, Ducloue et al.~\cite{Ducloue:2016pqr}). The model of Arleo et al.~\cite{Arleo:2014oha}, based on the calculation of the effects of parton coherent energy loss, gives a good description of backward-$y_{\rm {cms}}$ results and reproduces the data at forward $y_{\rm {cms}}$ fairly well. Finally, models including a contribution from the final state interaction of the c$\overline {\rm c}$ pair with the partonic/hadronic system created in the collision (Ferreiro~\cite{Ferreiro:2014bia}, Zhuang et al.~\cite{Chen:2016dke}) can also reproduce the trend observed in the data. In such a class of models nuclear shadowing is included, and is anyway the process that plays a dominant role in determining the values of the nuclear modification factors.

\begin{figure}
\begin{center}
\includegraphics[width=0.48\linewidth]{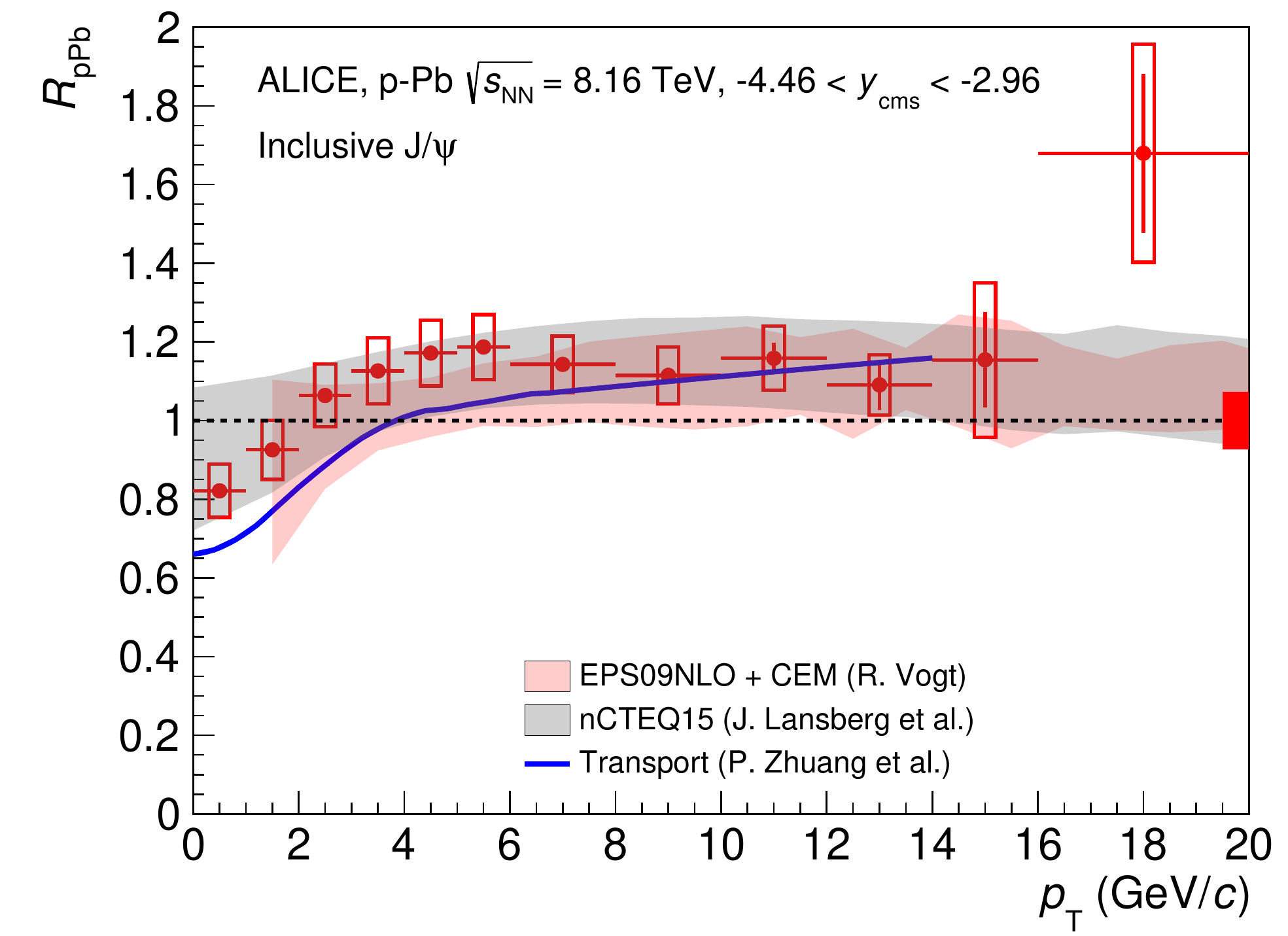}
\includegraphics[width=0.48\linewidth]{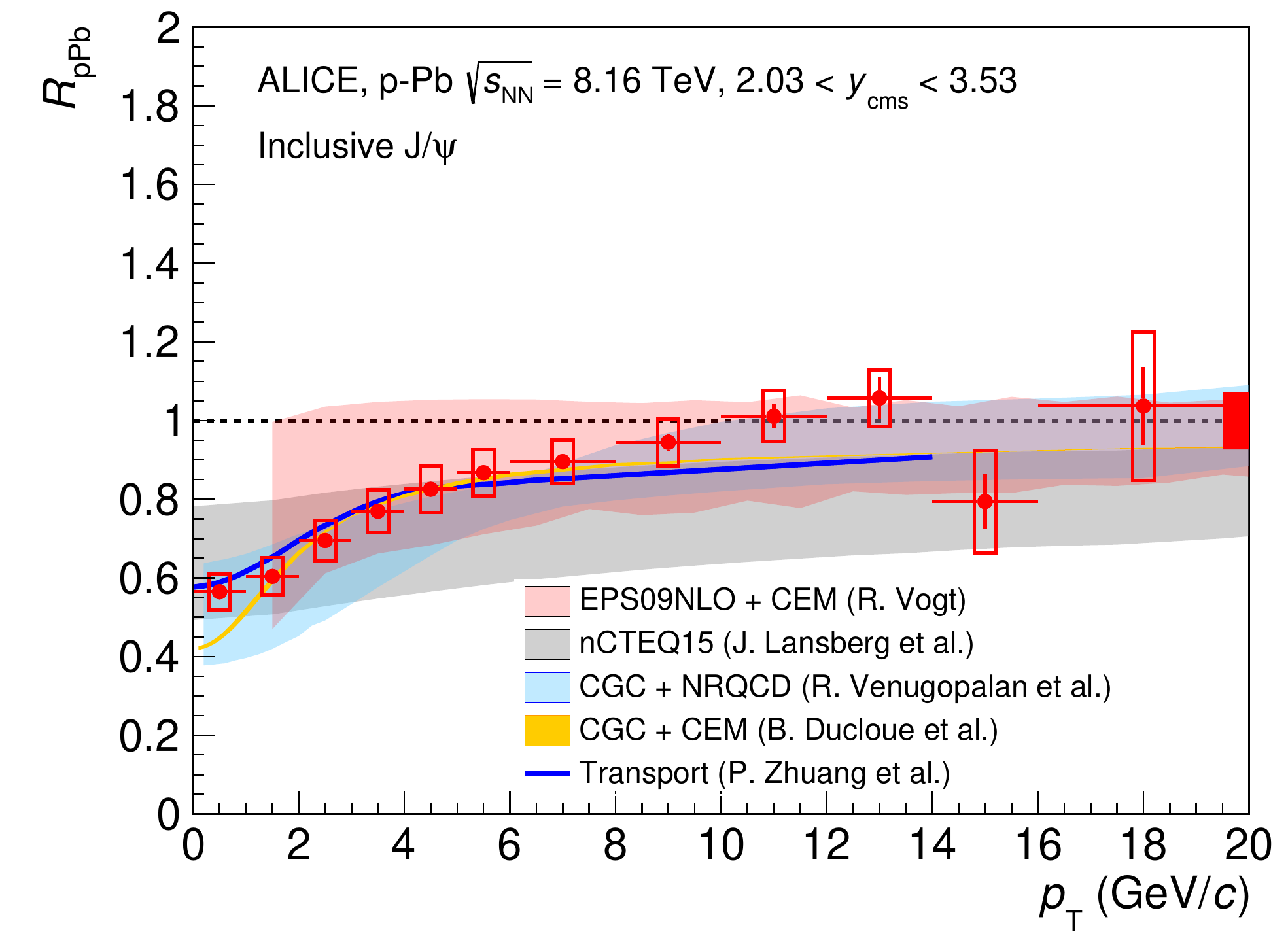}
\caption{Comparison of the ALICE results on the $p_{\rm T}$-dependence of the inclusive J/$\psi$ nuclear modification factors in \mbox{Pb--p} (left) and \mbox{p--Pb} (right) collisions at $\sqrt{s_{\rm NN}}=8.16$ TeV with model calculations~\cite{Albacete:2017qng,Lansberg:2016deg,Ma:2017rsu,Ducloue:2016pqr,Arleo:2014oha,Ferreiro:2014bia,Chen:2016dke} (see text for details). The horizontal bars on the experimental points correspond to the bin size. The vertical error bars represent the statistical uncertainties, the boxes around the points the uncorrelated systematic uncertainties and the filled box around unity the correlated uncertainties.}
\label{fig:rppbvsptmodels}
\end{center}
\end{figure}

Figure~\ref{fig:rppbvsptmodels} shows a comparison of the $p_{\rm T}$-dependence of $R_{\rm pPb}$ and $R_{\rm Pbp}$ with the calculations of the models discussed above. Thanks to the extended $p_{\rm T}$ range, these data explore a wide $x$ interval. At $y_{\rm cns}=2.78$ (centre of the forward-$y$ interval), the covered range for $0<p_{\rm T}<20$ GeV/$c$ is $2.3\cdot 10^{-5}< x < 1.5\cdot 10^{-4}$ while at $y_{\rm cms}=-3.71$ one has $1.5\cdot 10^{-2}< x < 10^{-1}$.
These values were calculated in the so-called $2\rightarrow 1$ approach, where the production channel is based on the gluon fusion process $gg\rightarrow {\rm J}/\psi$.
The agreement between data and models is rather good. It should be noted that for models that include uncertainty bands, such uncertainties are generally larger than those of the data, both as a function of $y_{\rm {cms}}$ and $p_{\rm T}$.

\begin{figure}
\begin{center}
\includegraphics[width=0.48\linewidth]{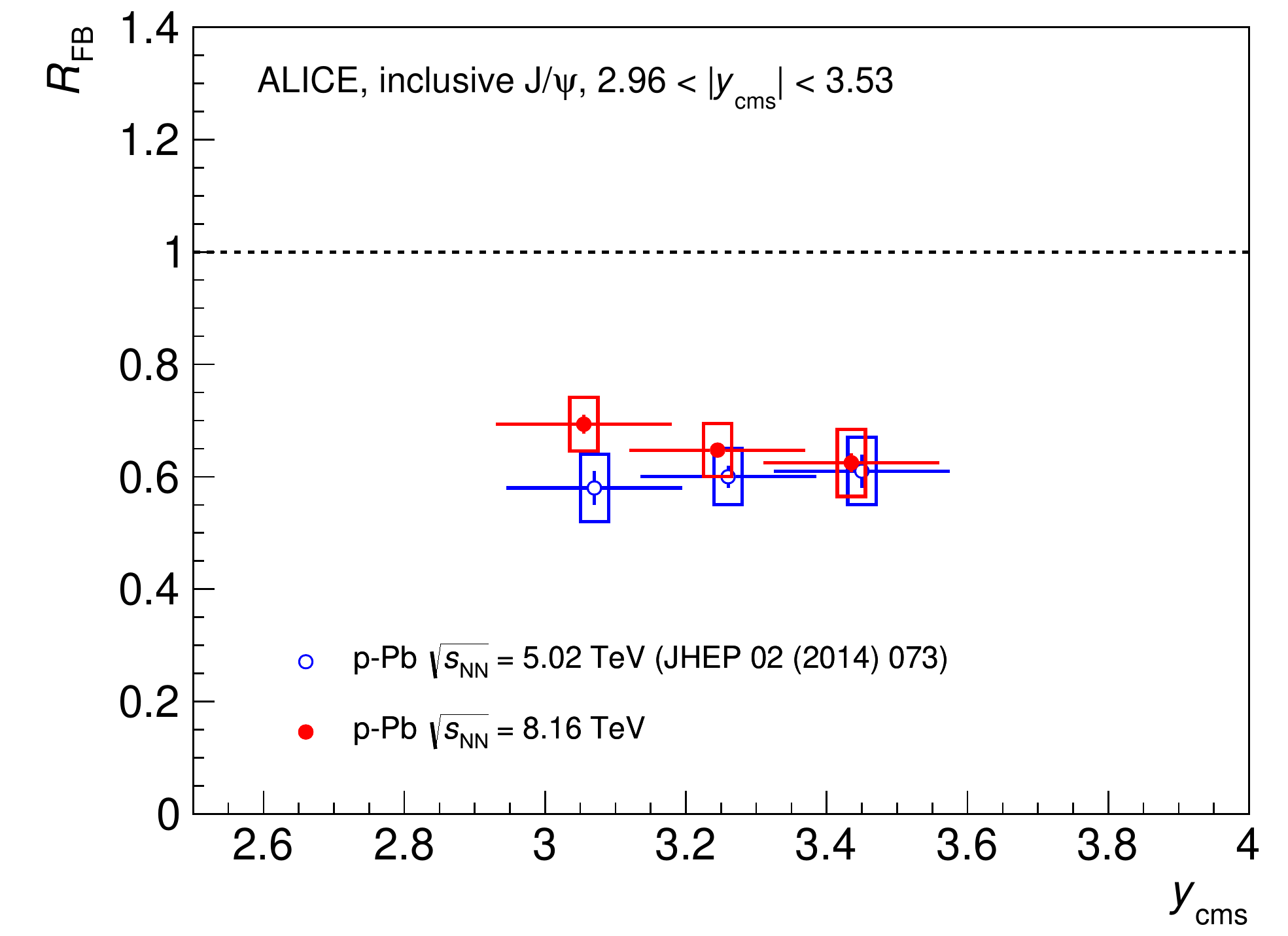}
\includegraphics[width=0.48\linewidth]{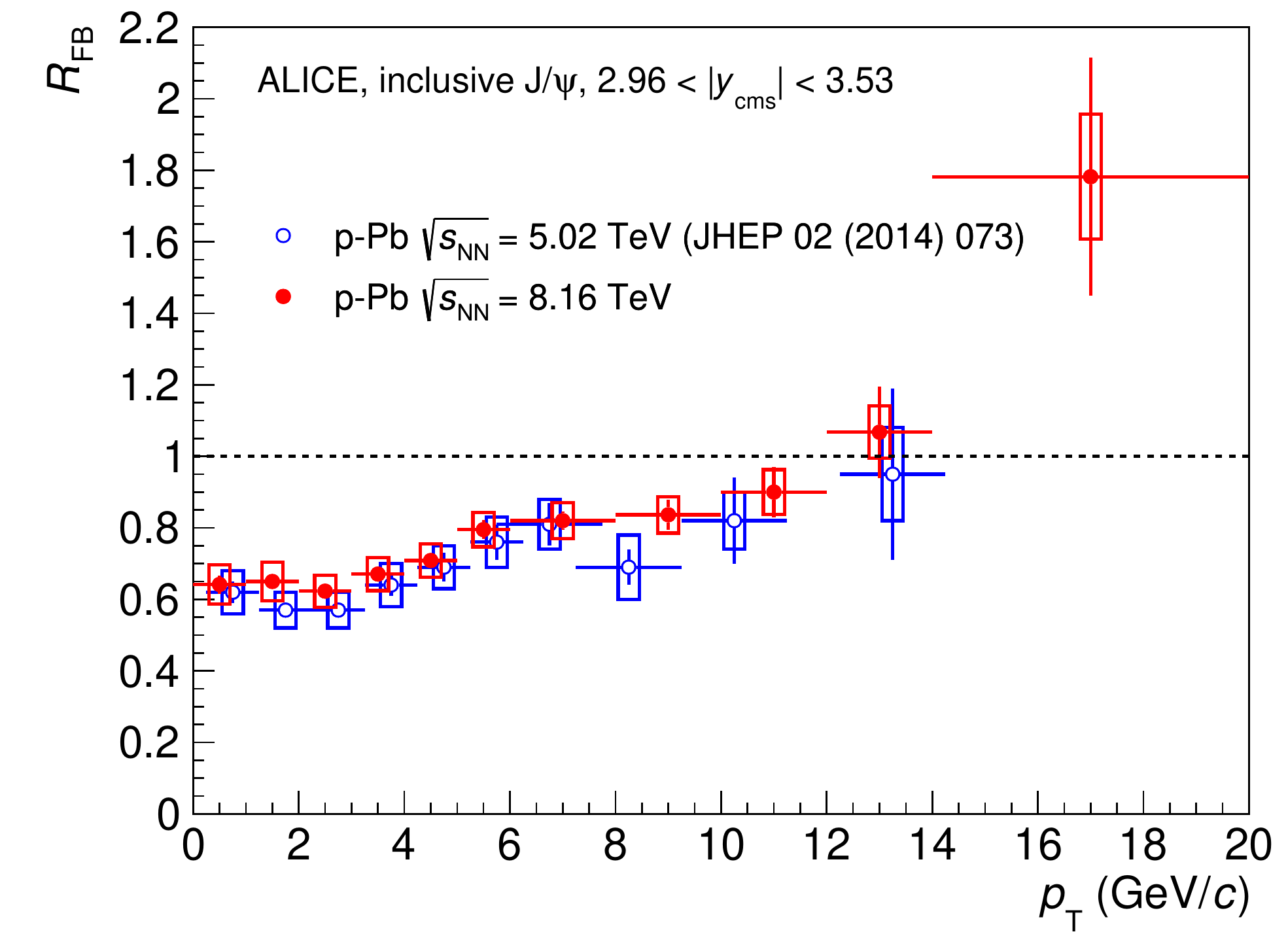}
\caption{The ratio $R_{\rm FB}$ between the inclusive J/$\psi$ nuclear modification factors, as a function of $y_{\rm cms}$ (left) and $p_{\rm T}$ (right), relative to $2.96<|y_{\rm {cms}}|<3.53$. The horizontal bars correspond to the bin size. The vertical error bars represent the statistical uncertainties, the boxes around the points the systematic uncertainties. The results are compared with those obtained at $\sqrt{s_{\rm NN}}=5.02$ TeV~\cite{Abelev:2013yxa}. The latter have been plotted at a slightly shifted $y_{\rm cms}$ and $p_{\rm T}$, for better visibility.}
\label{fig:RFB}
\end{center}
\end{figure}

By forming the ratio of the nuclear modification factors at forward and backward rapidity, it is possible to obtain a quantity, $R_{\rm FB}$, with smaller uncertainties, provided that the same absolute values of the $y_{\rm {cms}}$-ranges are chosen for the ratio. In this way, the reference pp cross section, and the related uncertainties, cancel out. $R_{\rm FB}$ is calculated in the rapidity range $2.96<|y_{\rm {cms}}|<3.53$, which is covered by both \mbox{p--Pb} and \mbox{Pb--p} samples. In Fig.~\ref{fig:RFB} the $y_{\rm {cms}}$- and $p_{\rm T}$-dependence of $R_{\rm FB}$ are shown, and compared with the corresponding results at $\sqrt{s_{\rm NN}}=5.02$ TeV~\cite{Abelev:2013yxa}. No appreciable dependence on $y_{\rm {cms}}$ can be seen, while  $R_{\rm FB}$ steadily increases as a function of $p_{\rm T}$, reaching unity at $p_{\rm T}\sim 12$ GeV/$c$. Results at $\sqrt{s_{\rm NN}}=8.16$ and 5.02 TeV are compatible within uncertainties.

\section{Conclusions}

Inclusive J/$\psi$ production in \mbox{p--Pb} collisions at $\sqrt{s_{\rm NN}}=8.16$ TeV was measured by ALICE, with about twice the integrated luminosity of the corresponding data sample at $\sqrt{s_{\rm NN}}=5.02$ TeV~\cite{Adam:2015iga}. Results on the cross sections and on the nuclear modification factors were shown, in six rapidity bins, for the p-going ($2.03<y_{\rm {cms}}<3.53$) and Pb-going ($-4.46<y_{\rm {cms}}<-2.96$) directions. The corresponding results as a function of transverse momentum were also shown, separately for the two $y_{\rm {cms}}$ regions, for $p_{\rm T}<20$ GeV/$c$. A suppression of the J/$\psi$ was observed at positive $y_{\rm {cms}}$, concentrated in the $p_{\rm T}\lesssim 5$ GeV/$c$ range. For negative $y_{\rm {cms}}$, an increasing trend in the nuclear modification factor is present at low $p_{\rm T}$, and the data are compatible with unity within 1.9$\sigma$ for $p_{\rm T}>4$ GeV/$c$. The ratios $R_{\rm FB}$ between forward- and backward-$y_{\rm {cms}}$ $R_{\rm pPb}$ in the region $2.96<|y_{\rm {cms}}|<3.53$ were also shown as a function of $y_{\rm {cms}}$ and $p_{\rm T}$. 
The results on the nuclear modification factors and on $R_{\rm FB}$ were found to be compatible with those obtained at $\sqrt{s_{\rm NN}}=5.02$ TeV. A good agreement is also observed when comparing ALICE and LHCb results at $\sqrt{s_{\rm NN}}=8.16$ TeV.
Finally, a comparison with several theory predictions shows that the results can be reproduced fairly well by calculations including various combinations of cold nuclear matter effects.

%
%

\newenvironment{acknowledgement}{\relax}{\relax}
\begin{acknowledgement}
\section*{Acknowledgements}

The ALICE Collaboration would like to thank all its engineers and technicians for their invaluable contributions to the construction of the experiment and the CERN accelerator teams for the outstanding performance of the LHC complex.
The ALICE Collaboration gratefully acknowledges the resources and support provided by all Grid centres and the Worldwide LHC Computing Grid (WLCG) collaboration.
The ALICE Collaboration acknowledges the following funding agencies for their support in building and running the ALICE detector:
A. I. Alikhanyan National Science Laboratory (Yerevan Physics Institute) Foundation (ANSL), State Committee of Science and World Federation of Scientists (WFS), Armenia;
Austrian Academy of Sciences and Nationalstiftung f\"{u}r Forschung, Technologie und Entwicklung, Austria;
Ministry of Communications and High Technologies, National Nuclear Research Center, Azerbaijan;
Conselho Nacional de Desenvolvimento Cient\'{\i}fico e Tecnol\'{o}gico (CNPq), Universidade Federal do Rio Grande do Sul (UFRGS), Financiadora de Estudos e Projetos (Finep) and Funda\c{c}\~{a}o de Amparo \`{a} Pesquisa do Estado de S\~{a}o Paulo (FAPESP), Brazil;
Ministry of Science \& Technology of China (MSTC), National Natural Science Foundation of China (NSFC) and Ministry of Education of China (MOEC) , China;
Ministry of Science and Education, Croatia;
Ministry of Education, Youth and Sports of the Czech Republic, Czech Republic;
The Danish Council for Independent Research | Natural Sciences, the Carlsberg Foundation and Danish National Research Foundation (DNRF), Denmark;
Helsinki Institute of Physics (HIP), Finland;
Commissariat \`{a} l'Energie Atomique (CEA) and Institut National de Physique Nucl\'{e}aire et de Physique des Particules (IN2P3) and Centre National de la Recherche Scientifique (CNRS), France;
Bundesministerium f\"{u}r Bildung, Wissenschaft, Forschung und Technologie (BMBF) and GSI Helmholtzzentrum f\"{u}r Schwerionenforschung GmbH, Germany;
General Secretariat for Research and Technology, Ministry of Education, Research and Religions, Greece;
National Research, Development and Innovation Office, Hungary;
Department of Atomic Energy Government of India (DAE), Department of Science and Technology, Government of India (DST), University Grants Commission, Government of India (UGC) and Council of Scientific and Industrial Research (CSIR), India;
Indonesian Institute of Science, Indonesia;
Centro Fermi - Museo Storico della Fisica e Centro Studi e Ricerche Enrico Fermi and Istituto Nazionale di Fisica Nucleare (INFN), Italy;
Institute for Innovative Science and Technology , Nagasaki Institute of Applied Science (IIST), Japan Society for the Promotion of Science (JSPS) KAKENHI and Japanese Ministry of Education, Culture, Sports, Science and Technology (MEXT), Japan;
Consejo Nacional de Ciencia (CONACYT) y Tecnolog\'{i}a, through Fondo de Cooperaci\'{o}n Internacional en Ciencia y Tecnolog\'{i}a (FONCICYT) and Direcci\'{o}n General de Asuntos del Personal Academico (DGAPA), Mexico;
Nederlandse Organisatie voor Wetenschappelijk Onderzoek (NWO), Netherlands;
The Research Council of Norway, Norway;
Commission on Science and Technology for Sustainable Development in the South (COMSATS), Pakistan;
Pontificia Universidad Cat\'{o}lica del Per\'{u}, Peru;
Ministry of Science and Higher Education and National Science Centre, Poland;
Korea Institute of Science and Technology Information and National Research Foundation of Korea (NRF), Republic of Korea;
Ministry of Education and Scientific Research, Institute of Atomic Physics and Romanian National Agency for Science, Technology and Innovation, Romania;
Joint Institute for Nuclear Research (JINR), Ministry of Education and Science of the Russian Federation and National Research Centre Kurchatov Institute, Russia;
Ministry of Education, Science, Research and Sport of the Slovak Republic, Slovakia;
National Research Foundation of South Africa, South Africa;
Centro de Aplicaciones Tecnol\'{o}gicas y Desarrollo Nuclear (CEADEN), Cubaenerg\'{\i}a, Cuba and Centro de Investigaciones Energ\'{e}ticas, Medioambientales y Tecnol\'{o}gicas (CIEMAT), Spain;
Swedish Research Council (VR) and Knut \& Alice Wallenberg Foundation (KAW), Sweden;
European Organization for Nuclear Research, Switzerland;
National Science and Technology Development Agency (NSDTA), Suranaree University of Technology (SUT) and Office of the Higher Education Commission under NRU project of Thailand, Thailand;
Turkish Atomic Energy Agency (TAEK), Turkey;
National Academy of  Sciences of Ukraine, Ukraine;
Science and Technology Facilities Council (STFC), United Kingdom;
National Science Foundation of the United States of America (NSF) and United States Department of Energy, Office of Nuclear Physics (DOE NP), United States of America.
\end{acknowledgement}

\bibliographystyle{utphys}   
\bibliography{JpsipPb8_paper_v6}

\newpage
\appendix
\section{The ALICE Collaboration}
\label{app:collab}

\begingroup
\small
\begin{flushleft}
S.~Acharya\Irefn{org138}\And 
F.T.-.~Acosta\Irefn{org22}\And 
D.~Adamov\'{a}\Irefn{org94}\And 
J.~Adolfsson\Irefn{org81}\And 
M.M.~Aggarwal\Irefn{org98}\And 
G.~Aglieri Rinella\Irefn{org36}\And 
M.~Agnello\Irefn{org33}\And 
N.~Agrawal\Irefn{org49}\And 
Z.~Ahammed\Irefn{org138}\And 
S.U.~Ahn\Irefn{org77}\And 
S.~Aiola\Irefn{org143}
% ...
\And
\renewcommand\labelenumi{\textsuperscript{\theenumi}~}

\section*{Affiliation notes}
\renewcommand\theenumi{\roman{enumi}}
\begin{Authlist}
\item \Adef{org*}Deceased
\item \Adef{orgI}Dipartimento DET del Politecnico di Torino, Turin, Italy
\item \Adef{orgII}M.V. Lomonosov Moscow State University, D.V. Skobeltsyn Institute of Nuclear, Physics, Moscow, Russia
\item \Adef{orgIII}Department of Applied Physics, Aligarh Muslim University, Aligarh, India
\item \Adef{orgIV}Institute of Theoretical Physics, University of Wroclaw, Poland
\end{Authlist}

\section*{Collaboration Institutes}
\renewcommand\theenumi{\arabic{enumi}~}
\begin{Authlist}
\item \Idef{org1}A.I. Alikhanyan National Science Laboratory (Yerevan Physics Institute) Foundation, Yerevan, Armenia
\item \Idef{org2}Benem\'{e}rita Universidad Aut\'{o}noma de Puebla, Puebla, Mexico
\item \Idef{org3}Bogolyubov Institute for Theoretical Physics, National Academy of Sciences of Ukraine, Kiev, Ukraine
\item \Idef{org4}Bose Institute, Department of Physics  and Centre for Astroparticle Physics and Space Science (CAPSS), Kolkata, India
\item \Idef{org5}Budker Institute for Nuclear Physics, Novosibirsk, Russia
\item \Idef{org6}California Polytechnic State University, San Luis Obispo, California, United States
\item \Idef{org7}Central China Normal University, Wuhan, China
\item \Idef{org8}Centre de Calcul de l'IN2P3, Villeurbanne, Lyon, France
\item \Idef{org9}Centro de Aplicaciones Tecnol\'{o}gicas y Desarrollo Nuclear (CEADEN), Havana, Cuba
\item \Idef{org10}Centro de Investigaci\'{o}n y de Estudios Avanzados (CINVESTAV), Mexico City and M\'{e}rida, Mexico
\item \Idef{org11}Centro Fermi - Museo Storico della Fisica e Centro Studi e Ricerche ``Enrico Fermi', Rome, Italy
\item \Idef{org12}Chicago State University, Chicago, Illinois, United States
\item \Idef{org13}China Institute of Atomic Energy, Beijing, China
\item \Idef{org14}Chonbuk National University, Jeonju, Republic of Korea
\item \Idef{org15}Comenius University Bratislava, Faculty of Mathematics, Physics and Informatics, Bratislava, Slovakia
\item \Idef{org16}COMSATS Institute of Information Technology (CIIT), Islamabad, Pakistan
\item \Idef{org17}Creighton University, Omaha, Nebraska, United States
\item \Idef{org18}Department of Physics, Aligarh Muslim University, Aligarh, India
\item \Idef{org19}Department of Physics, Ohio State University, Columbus, Ohio, United States
\item \Idef{org20}Department of Physics, Pusan National University, Pusan, Republic of Korea
\item \Idef{org21}Department of Physics, Sejong University, Seoul, Republic of Korea
\item \Idef{org22}Department of Physics, University of California, Berkeley, California, United States
\item \Idef{org23}Department of Physics, University of Oslo, Oslo, Norway
\item \Idef{org24}Department of Physics and Technology, University of Bergen, Bergen, Norway
\item \Idef{org25}Dipartimento di Fisica dell'Universit\`{a} 'La Sapienza' and Sezione INFN, Rome, Italy
\item \Idef{org26}Dipartimento di Fisica dell'Universit\`{a} and Sezione INFN, Cagliari, Italy
\item \Idef{org27}Dipartimento di Fisica dell'Universit\`{a} and Sezione INFN, Trieste, Italy
\item \Idef{org28}Dipartimento di Fisica dell'Universit\`{a} and Sezione INFN, Turin, Italy
\item \Idef{org29}Dipartimento di Fisica e Astronomia dell'Universit\`{a} and Sezione INFN, Bologna, Italy
\item \Idef{org30}Dipartimento di Fisica e Astronomia dell'Universit\`{a} and Sezione INFN, Catania, Italy
\item \Idef{org31}Dipartimento di Fisica e Astronomia dell'Universit\`{a} and Sezione INFN, Padova, Italy
\item \Idef{org32}Dipartimento di Fisica `E.R.~Caianiello' dell'Universit\`{a} and Gruppo Collegato INFN, Salerno, Italy
\item \Idef{org33}Dipartimento DISAT del Politecnico and Sezione INFN, Turin, Italy
\item \Idef{org34}Dipartimento di Scienze e Innovazione Tecnologica dell'Universit\`{a} del Piemonte Orientale and INFN Sezione di Torino, Alessandria, Italy
\item \Idef{org35}Dipartimento Interateneo di Fisica `M.~Merlin' and Sezione INFN, Bari, Italy
\item \Idef{org36}European Organization for Nuclear Research (CERN), Geneva, Switzerland
\item \Idef{org37}Faculty of Electrical Engineering, Mechanical Engineering and Naval Architecture, University of Split, Split, Croatia
\item \Idef{org38}Faculty of Engineering and Science, Western Norway University of Applied Sciences, Bergen, Norway
\item \Idef{org39}Faculty of Nuclear Sciences and Physical Engineering, Czech Technical University in Prague, Prague, Czech Republic
\item \Idef{org40}Faculty of Science, P.J.~\v{S}af\'{a}rik University, Ko\v{s}ice, Slovakia
\item \Idef{org41}Frankfurt Institute for Advanced Studies, Johann Wolfgang Goethe-Universit\"{a}t Frankfurt, Frankfurt, Germany
\item \Idef{org42}Gangneung-Wonju National University, Gangneung, Republic of Korea
\item \Idef{org43}Gauhati University, Department of Physics, Guwahati, India
\item \Idef{org44}Helmholtz-Institut f\"{u}r Strahlen- und Kernphysik, Rheinische Friedrich-Wilhelms-Universit\"{a}t Bonn, Bonn, Germany
\item \Idef{org45}Helsinki Institute of Physics (HIP), Helsinki, Finland
\item \Idef{org46}Hiroshima University, Hiroshima, Japan
\item \Idef{org47}Hochschule Worms, Zentrum  f\"{u}r Technologietransfer und Telekommunikation (ZTT), Worms, Germany
\item \Idef{org48}Horia Hulubei National Institute of Physics and Nuclear Engineering, Bucharest, Romania
\item \Idef{org49}Indian Institute of Technology Bombay (IIT), Mumbai, India
\item \Idef{org50}Indian Institute of Technology Indore, Indore, India
\item \Idef{org51}Indonesian Institute of Sciences, Jakarta, Indonesia
\item \Idef{org52}INFN, Laboratori Nazionali di Frascati, Frascati, Italy
\item \Idef{org53}INFN, Sezione di Bari, Bari, Italy
\item \Idef{org54}INFN, Sezione di Bologna, Bologna, Italy
\item \Idef{org55}INFN, Sezione di Cagliari, Cagliari, Italy
\item \Idef{org56}INFN, Sezione di Catania, Catania, Italy
\item \Idef{org57}INFN, Sezione di Padova, Padova, Italy
\item \Idef{org58}INFN, Sezione di Roma, Rome, Italy
\item \Idef{org59}INFN, Sezione di Torino, Turin, Italy
\item \Idef{org60}INFN, Sezione di Trieste, Trieste, Italy
\item \Idef{org61}Inha University, Incheon, Republic of Korea
\item \Idef{org62}Institut de Physique Nucl\'{e}aire d'Orsay (IPNO), Institut National de Physique Nucl\'{e}aire et de Physique des Particules (IN2P3/CNRS), Universit\'{e} de Paris-Sud, Universit\'{e} Paris-Saclay, Orsay, France
\item \Idef{org63}Institute for Nuclear Research, Academy of Sciences, Moscow, Russia
\item \Idef{org64}Institute for Subatomic Physics, Utrecht University/Nikhef, Utrecht, Netherlands
\item \Idef{org65}Institute for Theoretical and Experimental Physics, Moscow, Russia
\item \Idef{org66}Institute of Experimental Physics, Slovak Academy of Sciences, Ko\v{s}ice, Slovakia
\item \Idef{org67}Institute of Physics, Bhubaneswar, India
\item \Idef{org68}Institute of Physics of the Czech Academy of Sciences, Prague, Czech Republic
\item \Idef{org69}Institute of Space Science (ISS), Bucharest, Romania
\item \Idef{org70}Institut f\"{u}r Kernphysik, Johann Wolfgang Goethe-Universit\"{a}t Frankfurt, Frankfurt, Germany
\item \Idef{org71}Instituto de Ciencias Nucleares, Universidad Nacional Aut\'{o}noma de M\'{e}xico, Mexico City, Mexico
\item \Idef{org72}Instituto de F\'{i}sica, Universidade Federal do Rio Grande do Sul (UFRGS), Porto Alegre, Brazil
\item \Idef{org73}Instituto de F\'{\i}sica, Universidad Nacional Aut\'{o}noma de M\'{e}xico, Mexico City, Mexico
\item \Idef{org74}iThemba LABS, National Research Foundation, Somerset West, South Africa
\item \Idef{org75}Johann-Wolfgang-Goethe Universit\"{a}t Frankfurt Institut f\"{u}r Informatik, Fachbereich Informatik und Mathematik, Frankfurt, Germany
\item \Idef{org76}Joint Institute for Nuclear Research (JINR), Dubna, Russia
\item \Idef{org77}Korea Institute of Science and Technology Information, Daejeon, Republic of Korea
\item \Idef{org78}KTO Karatay University, Konya, Turkey
\item \Idef{org79}Laboratoire de Physique Subatomique et de Cosmologie, Universit\'{e} Grenoble-Alpes, CNRS-IN2P3, Grenoble, France
\item \Idef{org80}Lawrence Berkeley National Laboratory, Berkeley, California, United States
\item \Idef{org81}Lund University Department of Physics, Division of Particle Physics, Lund, Sweden
\item \Idef{org82}Nagasaki Institute of Applied Science, Nagasaki, Japan
\item \Idef{org83}Nara Women{'}s University (NWU), Nara, Japan
\item \Idef{org84}National and Kapodistrian University of Athens, School of Science, Department of Physics , Athens, Greece
\item \Idef{org85}National Centre for Nuclear Research, Warsaw, Poland
\item \Idef{org86}National Institute of Science Education and Research, HBNI, Jatni, India
\item \Idef{org87}National Nuclear Research Center, Baku, Azerbaijan
\item \Idef{org88}National Research Centre Kurchatov Institute, Moscow, Russia
\item \Idef{org89}Niels Bohr Institute, University of Copenhagen, Copenhagen, Denmark
\item \Idef{org90}Nikhef, National institute for subatomic physics, Amsterdam, Netherlands
\item \Idef{org91}NRC Kurchatov Institute IHEP, Protvino, Russia
\item \Idef{org92}NRNU Moscow Engineering Physics Institute, Moscow, Russia
\item \Idef{org93}Nuclear Physics Group, STFC Daresbury Laboratory, Daresbury, United Kingdom
\item \Idef{org94}Nuclear Physics Institute of the Czech Academy of Sciences, \v{R}e\v{z} u Prahy, Czech Republic
\item \Idef{org95}Oak Ridge National Laboratory, Oak Ridge, Tennessee, United States
\item \Idef{org96}Petersburg Nuclear Physics Institute, Gatchina, Russia
\item \Idef{org97}Physics department, Faculty of science, University of Zagreb, Zagreb, Croatia
\item \Idef{org98}Physics Department, Panjab University, Chandigarh, India
\item \Idef{org99}Physics Department, University of Jammu, Jammu, India
\item \Idef{org100}Physics Department, University of Rajasthan, Jaipur, India
\item \Idef{org101}Physikalisches Institut, Eberhard-Karls-Universit\"{a}t T\"{u}bingen, T\"{u}bingen, Germany
\item \Idef{org102}Physikalisches Institut, Ruprecht-Karls-Universit\"{a}t Heidelberg, Heidelberg, Germany
\item \Idef{org103}Physik Department, Technische Universit\"{a}t M\"{u}nchen, Munich, Germany
\item \Idef{org104}Research Division and ExtreMe Matter Institute EMMI, GSI Helmholtzzentrum f\"ur Schwerionenforschung GmbH, Darmstadt, Germany
\item \Idef{org105}Rudjer Bo\v{s}kovi\'{c} Institute, Zagreb, Croatia
\item \Idef{org106}Russian Federal Nuclear Center (VNIIEF), Sarov, Russia
\item \Idef{org107}Saha Institute of Nuclear Physics, Kolkata, India
\item \Idef{org108}School of Physics and Astronomy, University of Birmingham, Birmingham, United Kingdom
\item \Idef{org109}Secci\'{o}n F\'{\i}sica, Departamento de Ciencias, Pontificia Universidad Cat\'{o}lica del Per\'{u}, Lima, Peru
\item \Idef{org110}Shanghai Institute of Applied Physics, Shanghai, China
\item \Idef{org111}Stefan Meyer Institut f\"{u}r Subatomare Physik (SMI), Vienna, Austria
\item \Idef{org112}SUBATECH, IMT Atlantique, Universit\'{e} de Nantes, CNRS-IN2P3, Nantes, France
\item \Idef{org113}Suranaree University of Technology, Nakhon Ratchasima, Thailand
\item \Idef{org114}Technical University of Ko\v{s}ice, Ko\v{s}ice, Slovakia
\item \Idef{org115}Technische Universit\"{a}t M\"{u}nchen, Excellence Cluster 'Universe', Munich, Germany
\item \Idef{org116}The Henryk Niewodniczanski Institute of Nuclear Physics, Polish Academy of Sciences, Cracow, Poland
\item \Idef{org117}The University of Texas at Austin, Austin, Texas, United States
\item \Idef{org118}Universidad Aut\'{o}noma de Sinaloa, Culiac\'{a}n, Mexico
\item \Idef{org119}Universidade de S\~{a}o Paulo (USP), S\~{a}o Paulo, Brazil
\item \Idef{org120}Universidade Estadual de Campinas (UNICAMP), Campinas, Brazil
\item \Idef{org121}Universidade Federal do ABC, Santo Andre, Brazil
\item \Idef{org122}University College of Southeast Norway, Tonsberg, Norway
\item \Idef{org123}University of Cape Town, Cape Town, South Africa
\item \Idef{org124}University of Houston, Houston, Texas, United States
\item \Idef{org125}University of Jyv\"{a}skyl\"{a}, Jyv\"{a}skyl\"{a}, Finland
\item \Idef{org126}University of Liverpool, Department of Physics Oliver Lodge Laboratory , Liverpool, United Kingdom
\item \Idef{org127}University of Tennessee, Knoxville, Tennessee, United States
\item \Idef{org128}University of the Witwatersrand, Johannesburg, South Africa
\item \Idef{org129}University of Tokyo, Tokyo, Japan
\item \Idef{org130}University of Tsukuba, Tsukuba, Japan
\item \Idef{org131}Universit\'{e} Clermont Auvergne, CNRS/IN2P3, LPC, Clermont-Ferrand, France
\item \Idef{org132}Universit\'{e} de Lyon, Universit\'{e} Lyon 1, CNRS/IN2P3, IPN-Lyon, Villeurbanne, Lyon, France
\item \Idef{org133}Universit\'{e} de Strasbourg, CNRS, IPHC UMR 7178, F-67000 Strasbourg, France, Strasbourg, France
\item \Idef{org134} Universit\'{e} Paris-Saclay Centre d¿\'Etudes de Saclay (CEA), IRFU, Department de Physique Nucl\'{e}aire (DPhN), Saclay, France
\item \Idef{org135}Universit\`{a} degli Studi di Pavia, Pavia, Italy
\item \Idef{org136}Universit\`{a} di Brescia, Brescia, Italy
\item \Idef{org137}V.~Fock Institute for Physics, St. Petersburg State University, St. Petersburg, Russia
\item \Idef{org138}Variable Energy Cyclotron Centre, Kolkata, India
\item \Idef{org139}Warsaw University of Technology, Warsaw, Poland
\item \Idef{org140}Wayne State University, Detroit, Michigan, United States
\item \Idef{org141}Westf\"{a}lische Wilhelms-Universit\"{a}t M\"{u}nster, Institut f\"{u}r Kernphysik, M\"{u}nster, Germany
\item \Idef{org142}Wigner Research Centre for Physics, Hungarian Academy of Sciences, Budapest, Hungary
\item \Idef{org143}Yale University, New Haven, Connecticut, United States
\item \Idef{org144}Yonsei University, Seoul, Republic of Korea
\end{Authlist}
\endgroup
\end{document}